\documentclass[graybox]{svmult}

\usepackage{mathptmx}       
\usepackage{helvet}         
\usepackage{courier}        
\usepackage{type1cm}        
\usepackage{makeidx}         
\usepackage{graphicx}        
\usepackage{multicol}        
\usepackage[bottom]{footmisc}

\makeindex             

\begin{document}

\title*{Magnetic chemically peculiar stars}
\author{Markus Sch\"oller and Swetlana Hubrig}
\institute{Markus Sch\"oller \at European Southern Observatory, Karl-Schwarzschild-Str.~2, 85748 Garching, Germany, \email{mschoell@eso.org}
\and Swetlana Hubrig \at Leibniz-Institut f\"ur Astrophysik, An der Sternwarte~16, 14482 Potsdam, Germany \email{shubrig@aip.de}}
\maketitle

\abstract{
Chemically peculiar (CP) stars are main-sequence A and B stars with abnormally
strong or weak lines for certain elements.
They generally have magnetic fields and all observables tend to vary with the same
period.
Chemically peculiar stars provide a wealth of information; they are natural atomic
and magnetic laboratories.
After a brief historical overview, we discuss the general properties of the magnetic
fields in CP stars, describe the oblique rotator model, explain the dependence of
the magnetic field strength on the rotation, and concentrate at the end on HgMn stars.
}

\section{Introduction}
\label{sect:CP_intro}

\begin{table}[t]
\centering
\caption{
Different groups of chemically peculiar stars.
}
\label{tab:CPs}
\begin{tabular}{ccrcc}
\hline
\hline
\multicolumn{1}{c}{Peculiarity} &
\multicolumn{1}{c}{Spectral } &
\multicolumn{1}{c}{$T_{\rm eff}$} &
\multicolumn{1}{c}{magnetic} &
\multicolumn{1}{c}{spots} \\
\multicolumn{1}{c}{Type} &
\multicolumn{1}{c}{Type} &
\multicolumn{1}{c}{range} &
\multicolumn{1}{c}{} &
\multicolumn{1}{c}{} \\
\hline
He-strong & B1-B4 & $17\,000-21\,000$ & yes & yes \\
He-weak & B4-B8 & $13\,000-17\,000$ & yes & yes  \\
Si & B7-A0 & $9000-14\,000$ & yes & yes  \\
HgMn & B8-A0 & $10\,000-14\,000$ & yes? & yes!   \\
SrCrEu & A0-F0 & $7000-10\,000$ & yes & yes  \\
Am & A0-F0 & $7000-10\,000$ & yes? & no \\
\hline
\end{tabular}
\end{table}

Ap and Bp stars are
main sequence A and B stars, in the spectra of which lines of some elements are abnormally strong or weak
(e.g., Si, Sr, Cr, Eu, He, \dots).
They generally have magnetic fields that can be detected through observations of circular polarization in spectral lines.
Observables, such as the magnitudes in various photometric bands, the spectral line equivalent widths, and
the magnetic field, vary with the same period, which can range from half a day to several decades.
Abnormal line strengths correspond to element overabundances (by up to $5-$6\,dex with respect to the Sun) and are
confined to the stellar outer layers.
The class of chemically peculiar (CP) stars is roughly represented by three subclasses:
the magnetic Ap and Bp stars, the metallic-line Am stars, and the HgMn stars.
An overview of the different groups of CP stars can be
found in Table~\ref{tab:CPs}.

Chemically peculiar stars provide a wealth of information.
E.g., Castelli \& Hubrig (2004)
analyzed a spectrum of the HgMn star HD\,175640, observed with UVES at a spectral resolution of
$R\sim90\,000-100\,000$ and a spectral coverage of $3040-10\,000$\,\AA{}).
They used an ATLAS12 model atmosphere (Kurucz 1997)
with the SYNTHE code (Kurucz 1993) to model this spectrum.
They were able to obtain abundances for 49 ions, using 
200 lines for abundances of light elements, 
230 lines for abundances of ion group elements, and
130 lines for abundances of elements with $Z\ge31$.
They identified 80~Ti\,{\sc ii} emission lines, 40~Cr\,{\sc ii} emission lines,
and used 
100 lines to study the Mn\,{\sc ii} hyperfine structure,
140 lines for the Ga\,{\sc ii} isotopic structure,
15 lines for the Ba\,{\sc ii} hyperfine structure, and
30 lines for Hg\,{\sc ii} isotopic and hyperfine structure.
Still, there remained 170 unidentified absorption lines and 30 unidentified emission lines.

\begin{figure}[t]
\centering
\includegraphics[width=0.45\textwidth]{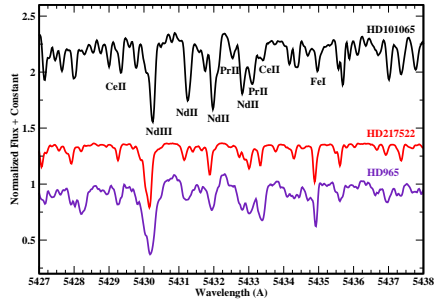}
\caption{
UVES observations of HD\,101065, HD\,217522, and HD\,965.
The magnetically insensitive Fe\,{\sc i} line at $\lambda\,5434$\,\AA{} is sharp
and has a similar width in all three spectra.
-- Credit: Hubrig et al.\ (2002).
}
\label{fig:CP_spectra}
\end{figure}

The difference between a non-peculiar star and a CP star can be striking.
Looking at the spectrum of Vega, in the region between 5000 and 6000\,\AA{} there
are only a few lines of Na\,{\sc i}, Mg\,{\sc i}, Si\,{\sc ii}, and Fe\,{\sc ii}. 
On the other hand, CP stars can have a dozen lines within a spectral range of
10\,\AA{}, which can be seen in Fig~\ref{fig:CP_spectra}.
The overabundances seen in CP stars are the result of 
selective diffusion of the different elements (Michaud 1970).
See also Chapter ``Diffusion and its manifestation in stellar atmospheres''.

\section{A brief historical overview}
\label{sect:history}

The first detection of a magnetic field in a star other than the Sun was achieved in CS\,Vir
by Babcock (1947).
He essentially determined the longitudinal magnetic field in this star.
Today, mean longitudinal magnetic field measurements throughout the variation period
have been obtained for no more than 100 stars.
The resolution of magnetically split lines requires a strong enough magnetic field and sufficiently slow rotation.
Resolved magnetically split lines were first discovered in Babcock's star, HD\,215441 (Babcock 1960),
for which he measured a mean magnetic field modulus of $\left<B\right> \sim 34$\,kG, and which is
the strongest magnetic field modulus measured in an Ap star to date.
In 1987, twelve stars with magnetically resolved lines were known, only four of those were
studied throughout their variation period.
In 2001, 44 stars with magnetically resolved lines were known,
24 of those were studied throughout their variation period
(Mathys et al.\ 1997; Mathys et al., {\sl in preparation}).
First systematic determinations of the 
crossover effect and the mean quadratic magnetic field were
published by Mathys (1995a,b).
A full phase coverage was achieved for about two dozen stars.
The bulk of the published material on broad-band linear polarization (BBLP) was
gathered by Leroy between 1990 and 1995 (Leroy 1995, and references therein).
Variations in BBLP were well studied for about 15 stars.
See Chapter ``Magnetic fields'' on a detailed discussion of stellar magnetic fields.

\section{General properties of magnetic fields in Ap stars}
\label{sect:properties}

The strongest magnetic fields tend to be found in more massive stars.
They are also found only in fast-rotating stars (Hubrig et al.\ 2000).
All stars with rotation periods exceeding 1000\,days have magnetic fields below 6.5\,kG.
From the finding that the longitudinal magnetic field averaged over the stellar disk is not zero,
one can directly conclude that the magnetic field needs to be organized on a larger scale,
i.e.\ it is a dipole or a superposition of a dipole and a quadrupole. 
The circular polarization from tangled, solar-like magnetic fields mostly cancels out in a disk integration.
The magnetic field of Ap stars thus has a significant dipole-like component.
For a dipole, the ratio between the longitudinal magnetic field and the magnetic field modulus
$\left<B_{\rm z}\right>/\left<B\right>$ is 0.3, for a quadrupole it is 0.05.
If toroidal or higher-order multipolar components were sufficient to account
for the observed longitudinal magnetic field, these  would induce strong distortions
of the spectral line profiles in Stokes~$I$, i.e.\ in integral light, which we do not see.

The magnetic field covers the whole stellar surface homogeneously, i.e.\ 
the distribution of the field strength over the star is fairly narrow.
Evidence for this comes from the fact that the magnetic field is observed at all phases,
the continuum is reached between the split components of resolved lines, and that
the resolved magnetically split components are rather narrow (Mathys et al.\ 1997).

Magnetic fields have severe effects on the structure of stellar outer layers.
They are responsible for magnetically controlled winds and
elemental abundance stratification.
Evidence for abnormal atmospheric structure comes from the fact that
profiles of hydrogen Balmer lines in cool Ap stars can not be fitted by conventional models
(Ryabchikova et al.\ 2002).
This has also a potential impact on the longitudinal magnetic field determination by Balmer line polarimetry.
The core-wing anomaly (Cowley et al.\ 2001) of the hydrogen Balmer lines leads to the impossibility of fitting the
Balmer lines with one effective temperature.
E.g., to fit the H$\beta$ line in HD\,965, one needs to assume 
$T_{\rm eff} = 5500$\,K for the core of the line and 
$T_{\rm eff} = 7000$\,K for the wings.

\section{The oblique rotator and the geometric structure of the magnetic field}
\label{sect:geometry}

The magnetic field is not symmetric with respect to the stellar rotation axis.
Other surface features, e.g.\ the abundance distribution, are determined by the magnetic field.
Observed variations result from changing aspects of the visible hemisphere as the star rotates.
Thus, the variation period is the rotation period of the star.
No intrinsic variations of the magnetic field have been observed in Ap stars over timescales of decades.

In early models of the magnetic field, a
quasi-sinusoidal variation of the longitudinal magnetic field was assumed.
In the simplest model, a dipole centered at the star's center and with an axis
inclined with respect to the stellar rotation axis, was employed.
From stars with magnetically resolved lines, it can be seen that
the mean magnetic field modulus generally has one maximum and one minimum per rotation period,
even for stars with a reversing longitudinal magnetic field (Mathys et al.\ 1997).
From these observations, a centered dipole can be ruled out.
Alternative models include a
dipole that is offset along its axis (parameters: $i$, $\beta$, $B_{\rm d}$, $a$), or
a collinear dipole plus a  quadrupole (parameters: $i$, $\beta$, $B_{\rm d}$, $B_{\rm q}$),
with $i$ the inclination angle of the star with respect to the line of sight,
$\beta$ the inclination angle of the magnetic field with respect to $i$,
$B_{\rm d}$ the strength of the dipole,
$B_{\rm q}$ the strength of the quadrupole,
and $a$ the offset of the dipole with respect to the star's center.
The models have to make a good match with four observables:
the maximum and the minimum of both the longitudinal magnetic field and the magnetic field modulus.
Both models are equivalent to first order.

Additional constraints on the magnetic field geometry can come from the
cross\-over and the  mean quadratic magnetic field.
A collinear dipole plus a quadrupole and an octupole give good first approximations in many cases
(Landstreet \& Mathys 2000).
The dipole primarily accounts for the longitudinal magnetic field, the
quadrupole gives the field strength contrast between the poles, and the
octupole is responsible for the equator-to-pole field strength contrast.
Asymmetric variation curves can be determined from some magnetic field moments.
They exist, if the magnetic field is not symmetric about an axis passing
through the center of the star (Mathys 1993) and can be described with a
generalized multipolar model (Bagnulo et al.\ 2000, and references therein).
The input observables for these models are all available observables of the magnetic field:
$\left<B_{\rm z}\right>$,
$\left<x B_{\rm z}\right>$,
$\sqrt{\left<B^2\right>+\left<B_{\rm z}^2\right>}$,
$\left<B\right>$, and the BBLP.
A $\chi^2$ minimization between the predicted and the observed values
of the observables at phases distributed throughout the rotation period will determine the final model
for the geometric structure of the magnetic field.

Ultimately, a direct inversion of the line profiles recorded in all four Stokes parameters
will allow one to derive magnetic field maps without a priori assumptions.
Since the inversion is an ill-posed problem, a regularization condition is needed.
This is achieved with the magnetic Doppler imaging technique
(Piskunov \& Kochukhov 2002).
It is very demanding in terms of the signal-to-noise ratio in the data,
spectral resolution, and phase coverage.
So far, these inversions are restricted to a few individual stars
(e.g., L\"uftinger et al.\ 2010;
Kochukhov et al.\ 2004).

\section{Field strength distribution and rotation}
\label{sect:distribution}

The mean longitudinal magnetic field distribution extends all the way down to the detection limit
of 100\,G or less (Landstreet 1982).
The rms mean longitudinal magnetic field averaged over a stellar rotation period
is of the order of 300\,G for ``classical'' Ap stars,
and larger ($\sim1$\,kG) for hotter He weak and He strong Bp stars (Landstreet 1982).

The mean magnetic field modulus much better characterizes the intrinsic stellar
magnetic field than the mean longitudinal magnetic field, which is much more dependent on the geometry of the observation.
Most Ap stars with magnetically resolved lines have a mean magnetic field modulus
(averaged over the stellar rotation period) comprised between 3 and 9\,kG.
But there is a lower cutoff of the distribution at 2.8\,kG.
One expects to be able to resolve lines down to 1.7\,kG or lower
at some rotation phases of some stars,
but only for one target it is observed down to 2.2\,kG.
The lower limit of the magnetic field distribution is roughly temperature independent;
hotter stars may have stronger magnetic fields than cooler stars (Mathys et al.\ 1997).

Ap star variation periods span five orders of magnitude.
Until recently, there seemed to be no systematic differences between short and long period stars.
A confirmation that very long periods are indeed rotation periods has been brought by BBLP
(Leroy  et al.\ 1994).
The systematic study of Ap stars with resolved magnetically split lines
has doubled the number of known stars with $P > 30$\,days.
The distribution of periods longer than 1\,year is compatible with an equipartition on a logarithmic scale.
No star with $P > 150$\,d has a mean magnetic field modulus exceeding 7.5\,kG.
More than 50\% of the stars with resolved lines and shorter periods
have a magnetic field modulus above this value (Mathys et al.\ 1997).
In the collinear dipole plus quadrupole and octupole model,
the angle between the magnetic and rotation axis $\beta$ is generally smaller
than 20$^{\circ}$ for stars with $P > 30$\,d, unlike for short period magnetic Ap stars,
for which this angle is usually large (Landstreet \& Mathys 2000).

\section{HgMn stars}

HgMn stars are chemically peculiar stars with spectral type B8 to A0 and
$T_{\rm eff}=10\,000-14\,000$\,K.
They show extreme overabundance of Hg (up to 6\,dex) and/or Mn (up to 3\,dex).
They display the most obvious departures from abundances expected within the context of nucleosynthesis
(Cowley \& Aikman 1975).
More than 150 HgMn stars are known,
many of which are found in young associations (Sco-Cen, Orion OB1).
They are among the most slowly rotating stars on the upper main sequence and have exceptionally stable atmospheres with
an average rotational velocity of $\left<v \sin i\right>=29$\,km/s, which leads to
extremely sharp-lined spectra.
They are the best suited targets to study isotopic and hyperfine structure.
More than 2/3 of the HgMn stars belong to SB systems with a prevalence of $P_{\rm orb}\approx3-20$\,d.
Many HgMn stars are in multiple systems.
The spectrum variability seen in HgMn stars is due to the presence of chemical spots.
They do not have strong large-scale organized magnetic fields, but tangled magnetic fields are possible.
They do not have enhanced strengths of rare earth elements, but of the heavy elements
W, Re, Os, Ir, Pt, Au, Hg, Tl, Pb, and Bi, which makes them a natural laboratory for the study of heavy elements.
They also show anomalous isotopic abundances for the elements He, Hg, Pt, Tl, Pb, and Ca.

\begin{figure}[t]
\centering
\includegraphics[width=0.115\textwidth, angle=0]{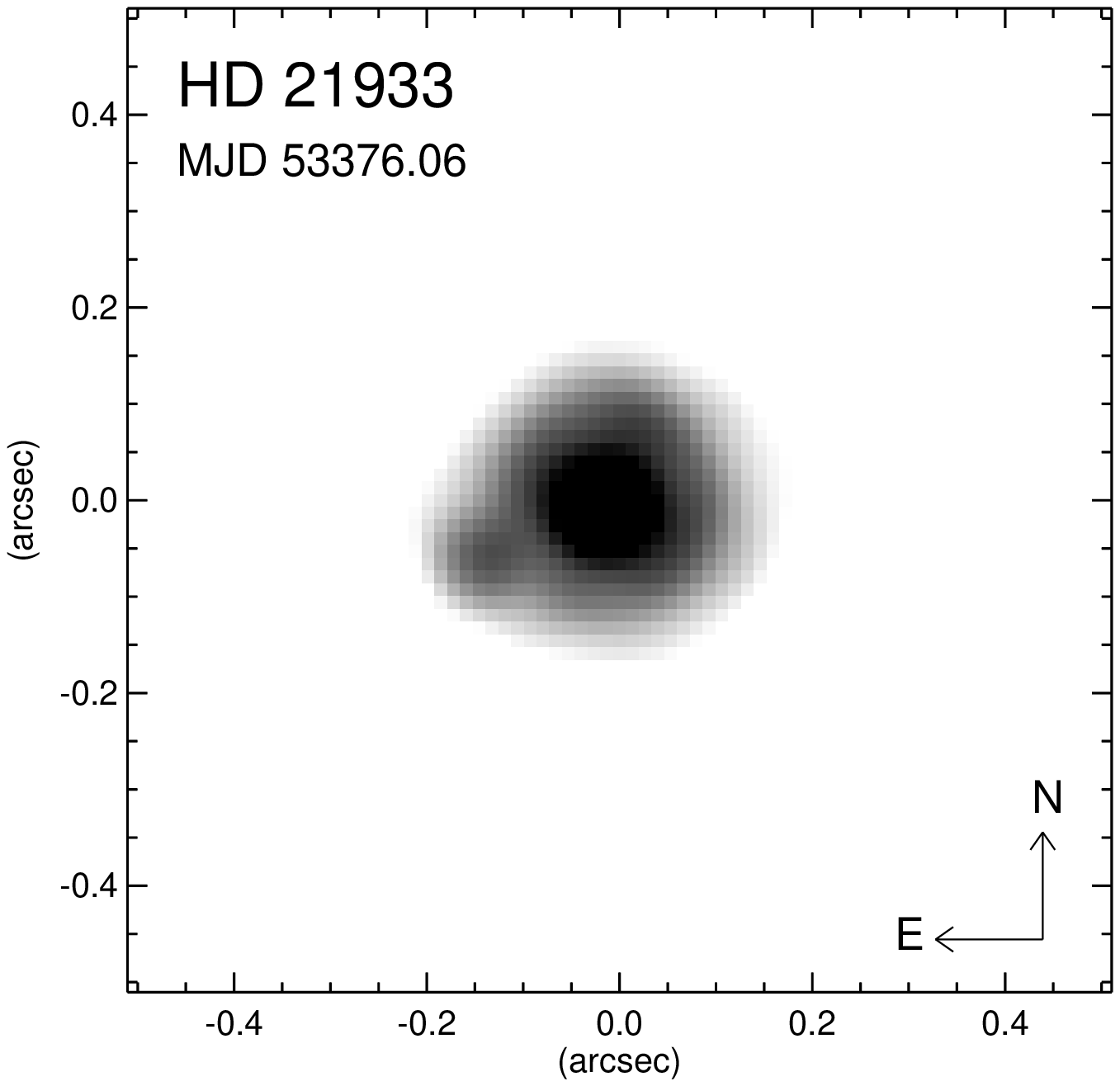}
\includegraphics[width=0.115\textwidth, angle=0]{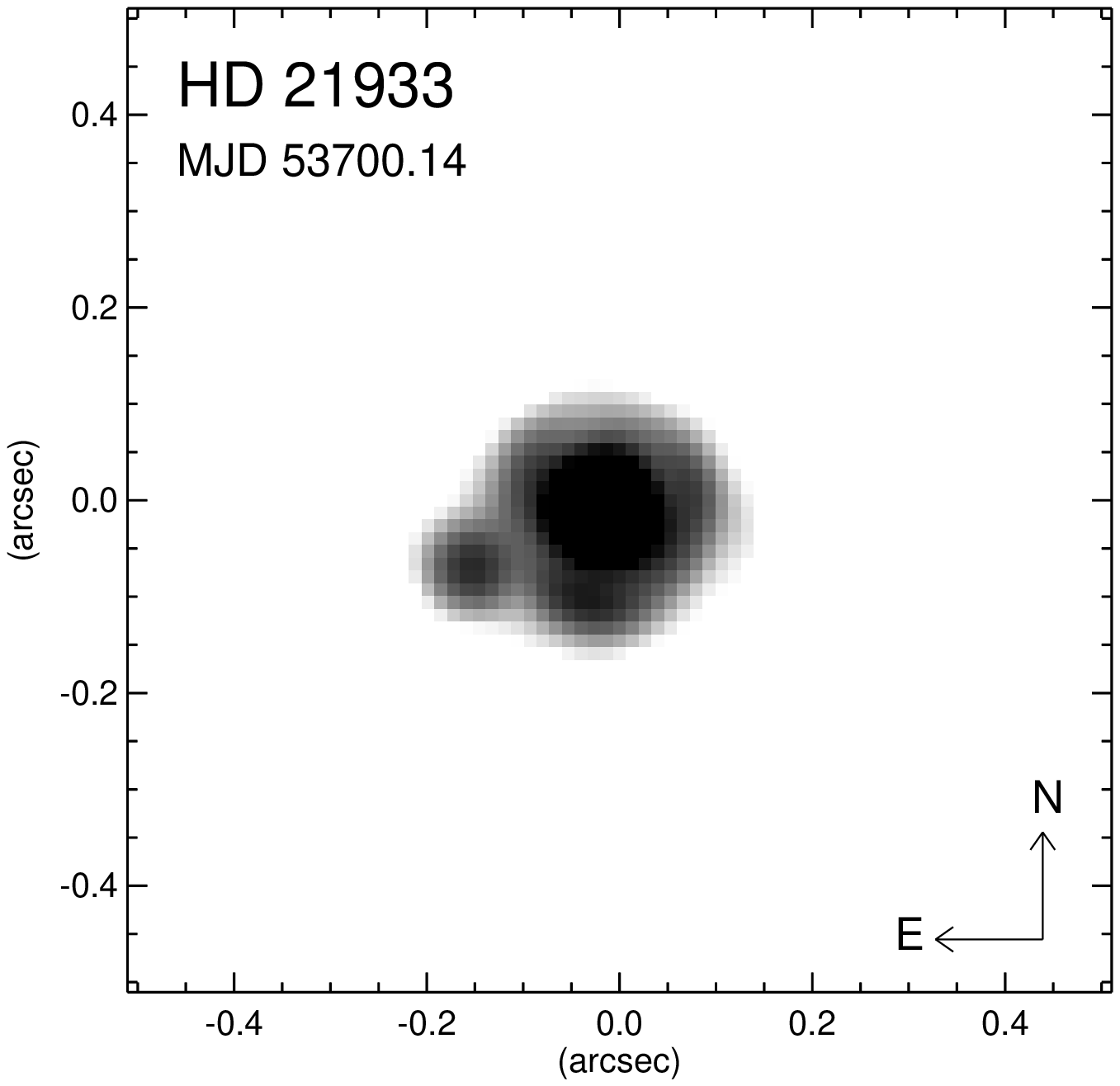}
\includegraphics[width=0.115\textwidth, angle=0]{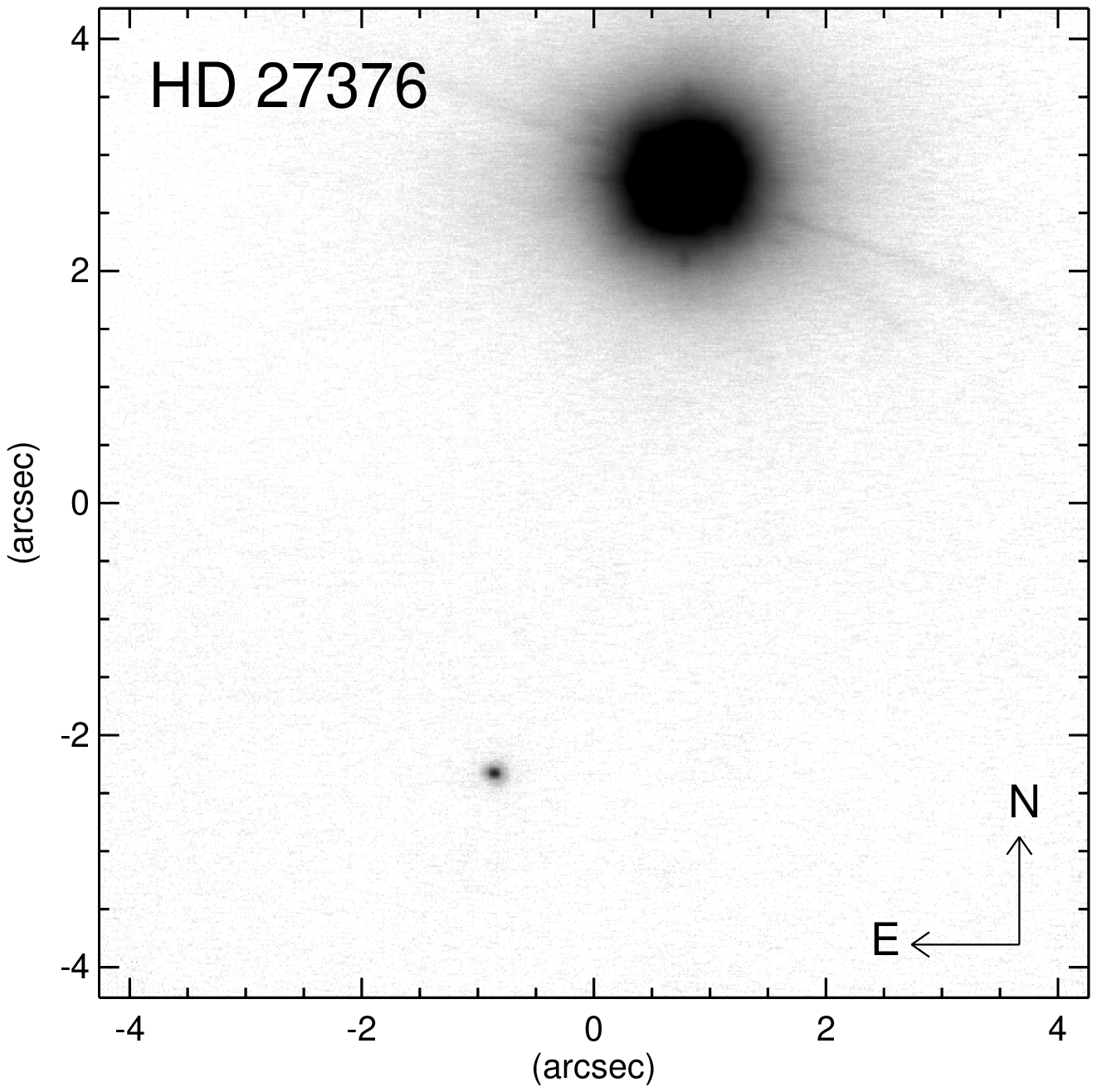}
\includegraphics[width=0.115\textwidth, angle=0]{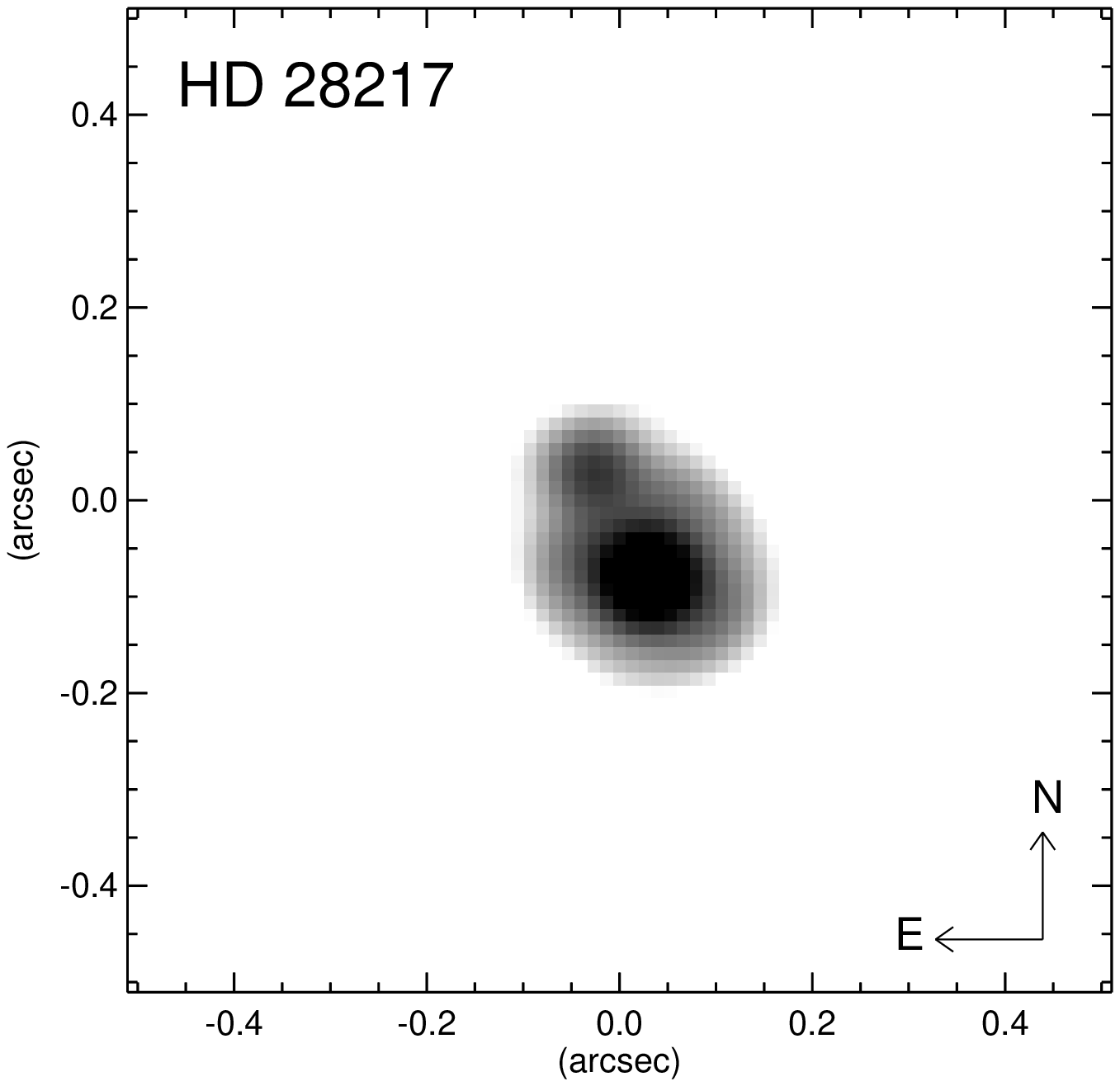}
\includegraphics[width=0.115\textwidth, angle=0]{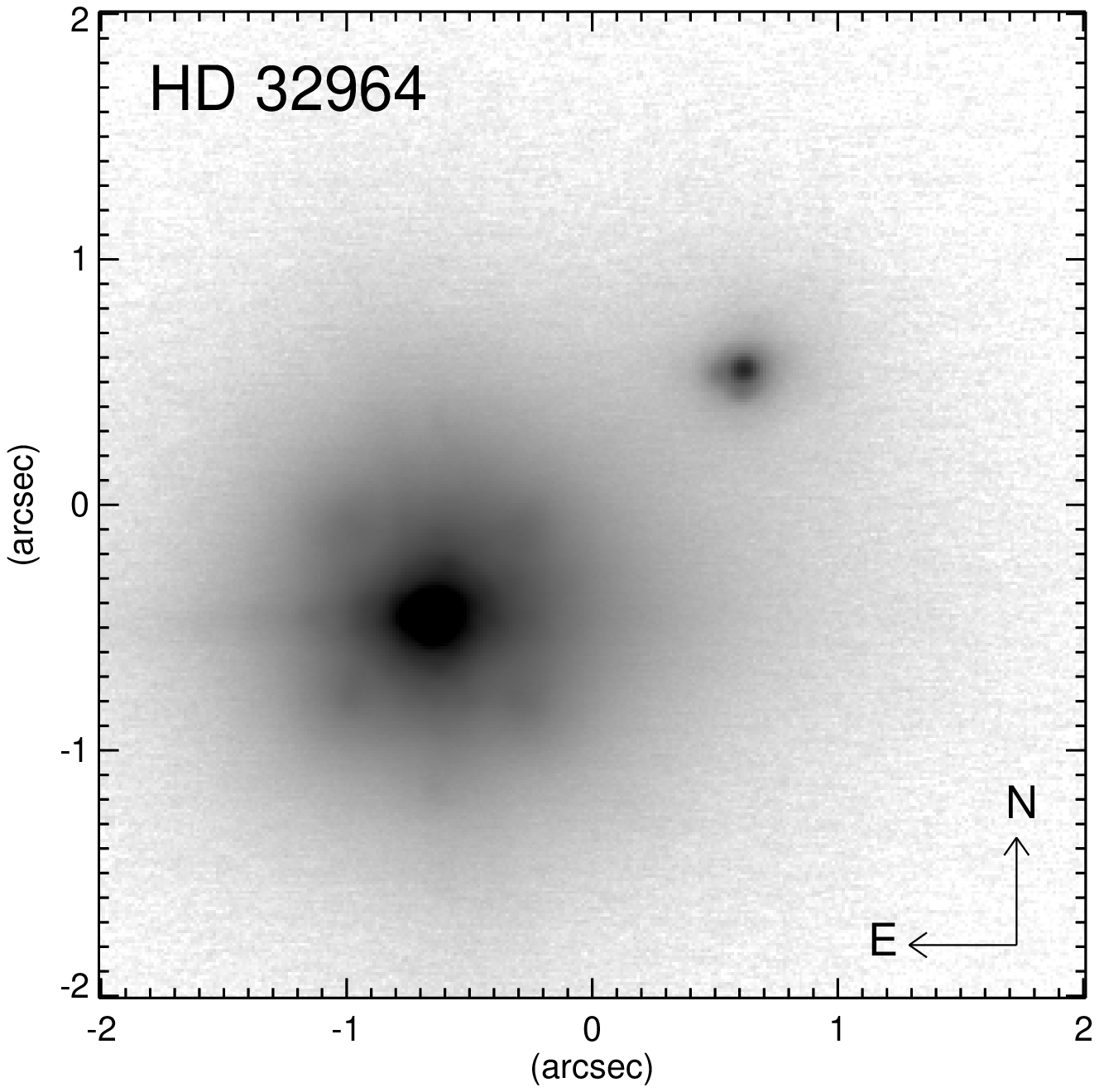}
\includegraphics[width=0.115\textwidth, angle=0]{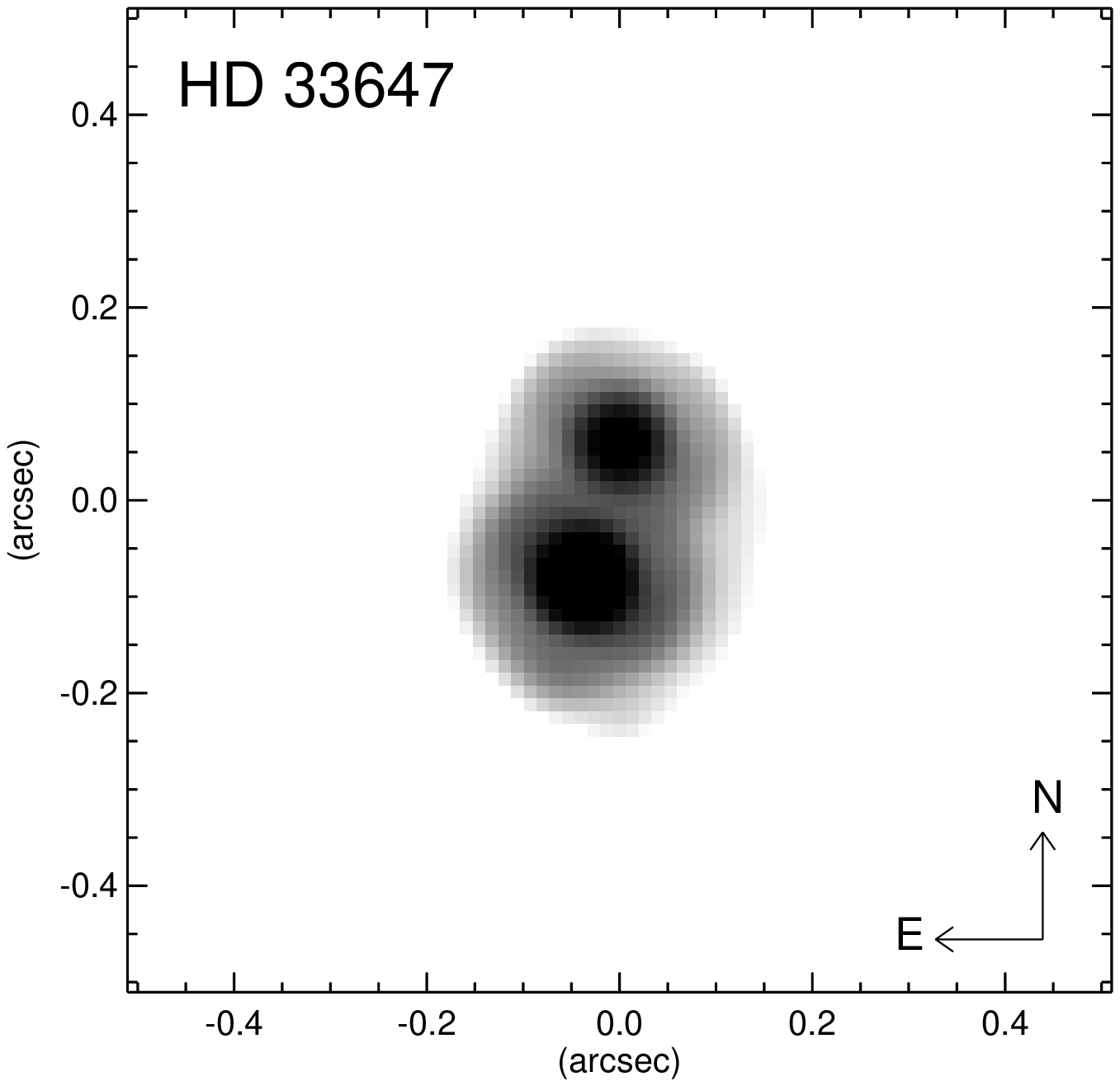}
\includegraphics[width=0.115\textwidth, angle=0]{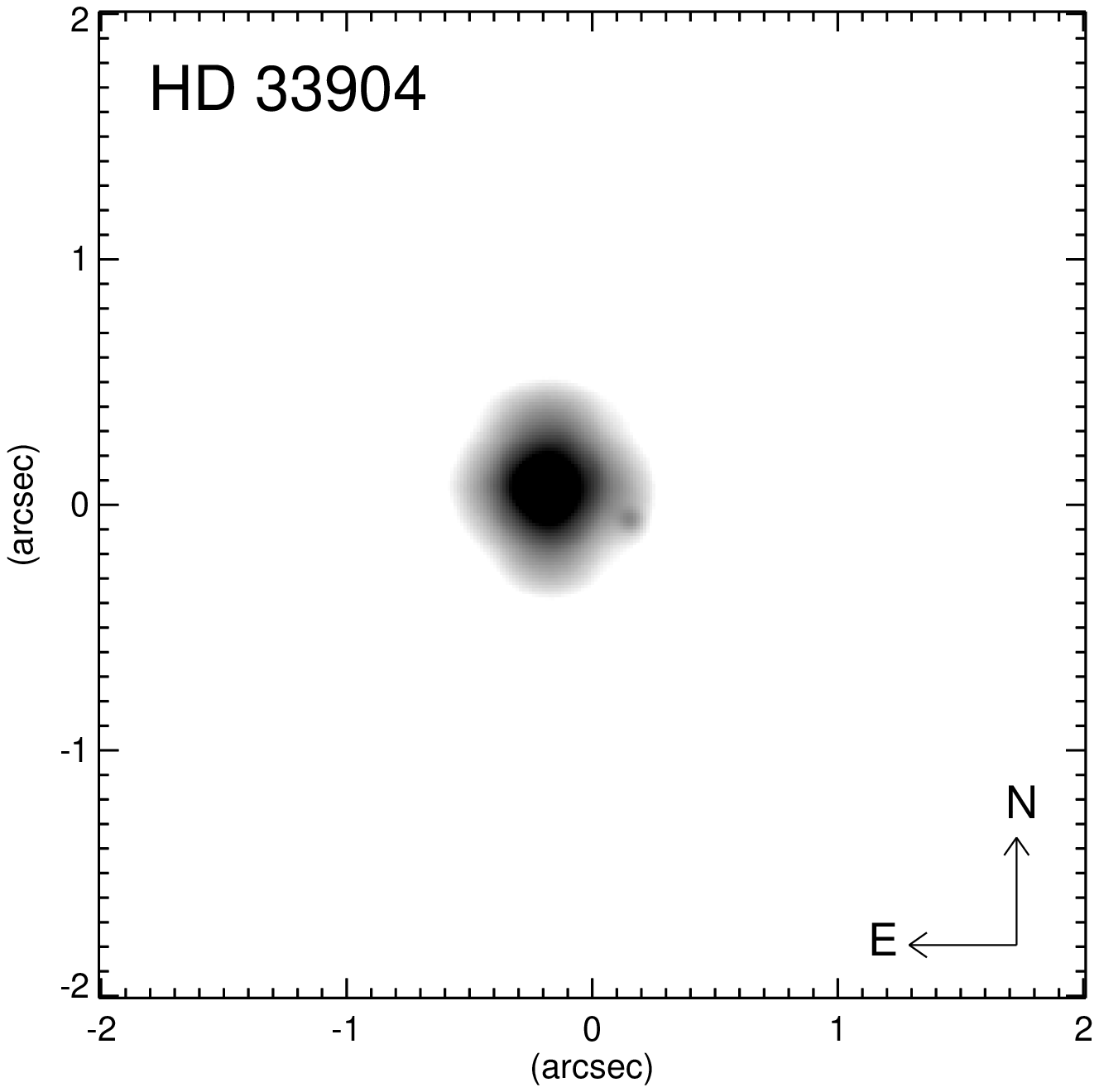}
\includegraphics[width=0.115\textwidth, angle=0]{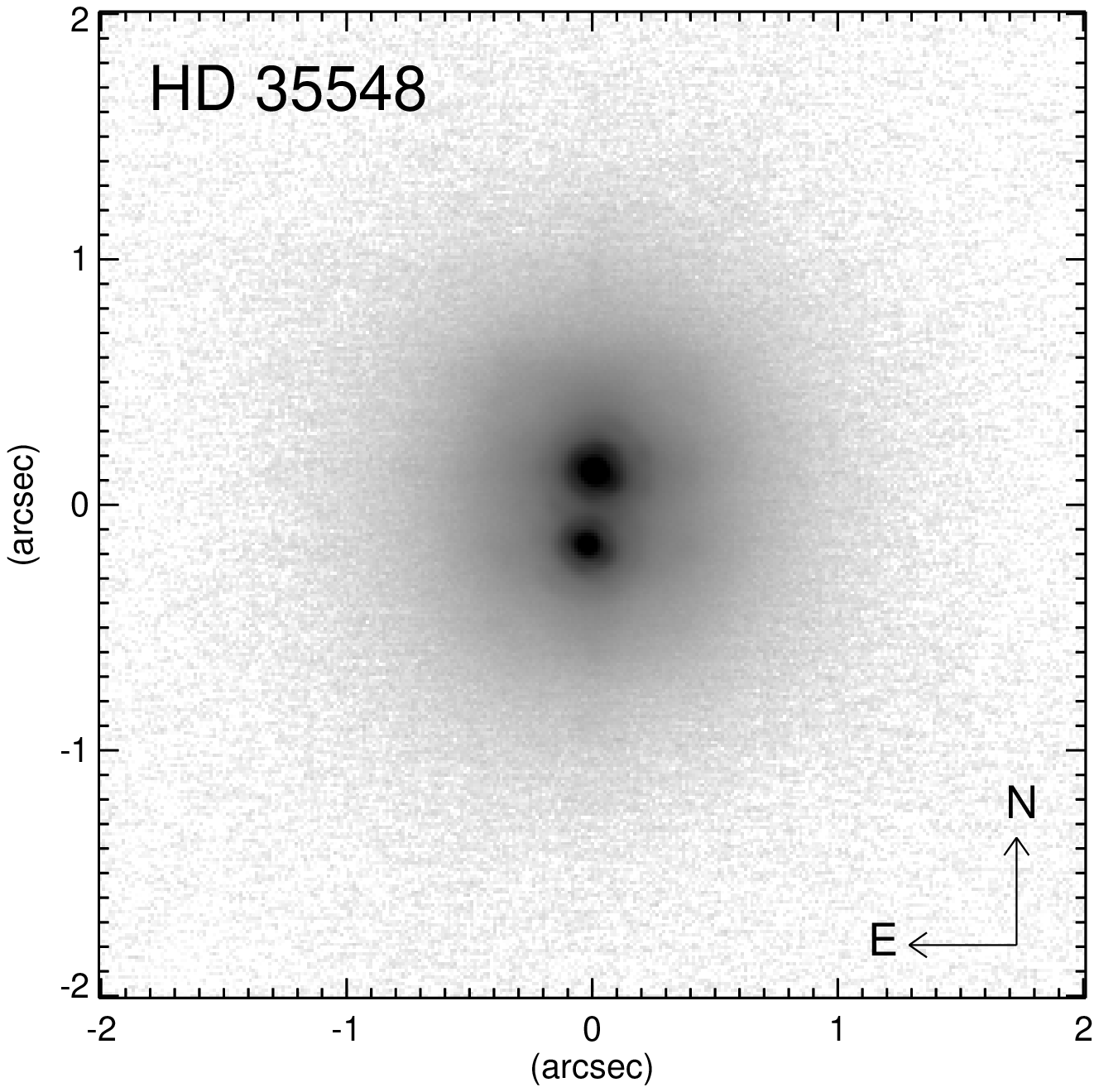}
\includegraphics[width=0.115\textwidth, angle=0]{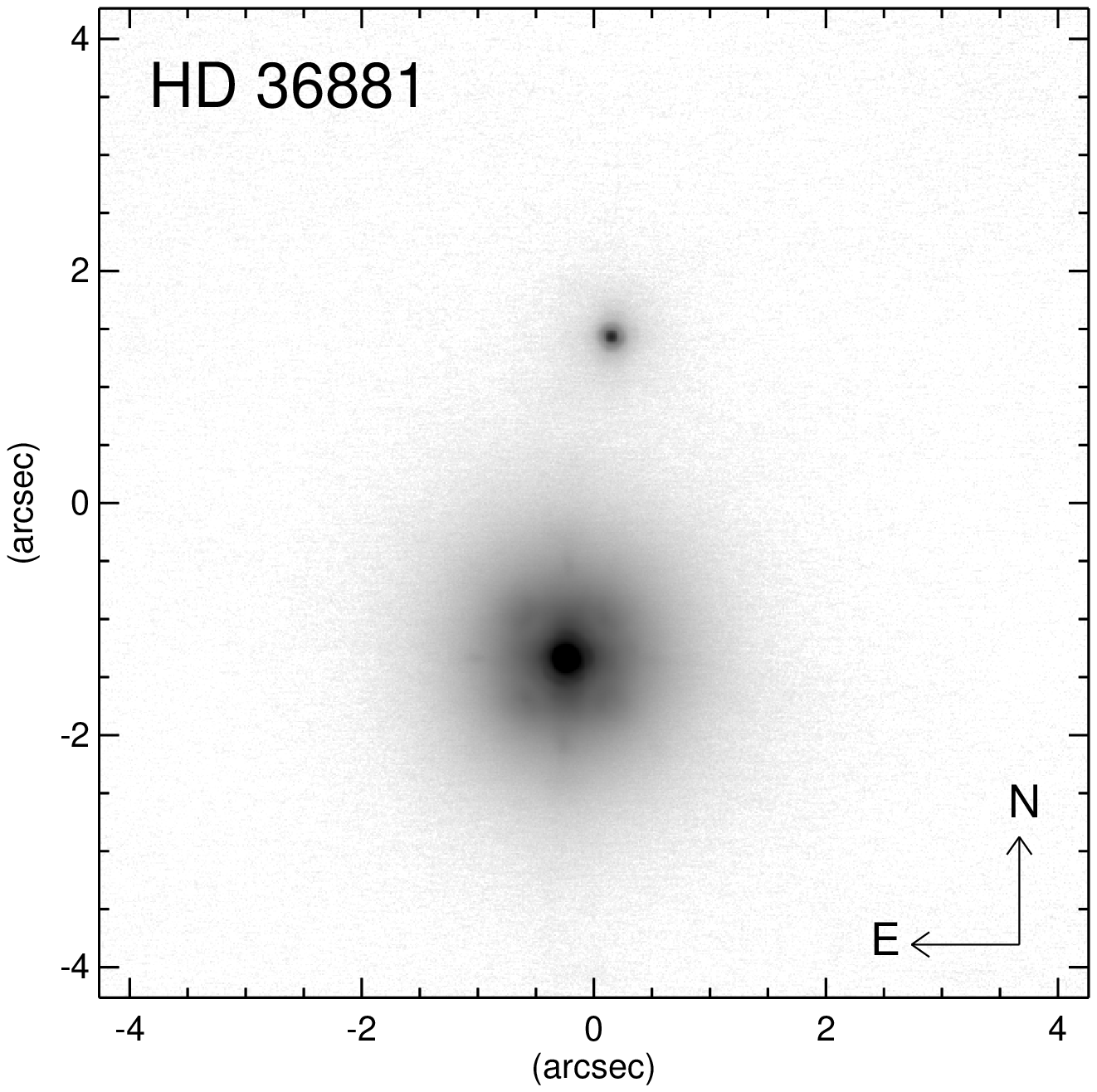}
\includegraphics[width=0.115\textwidth, angle=0]{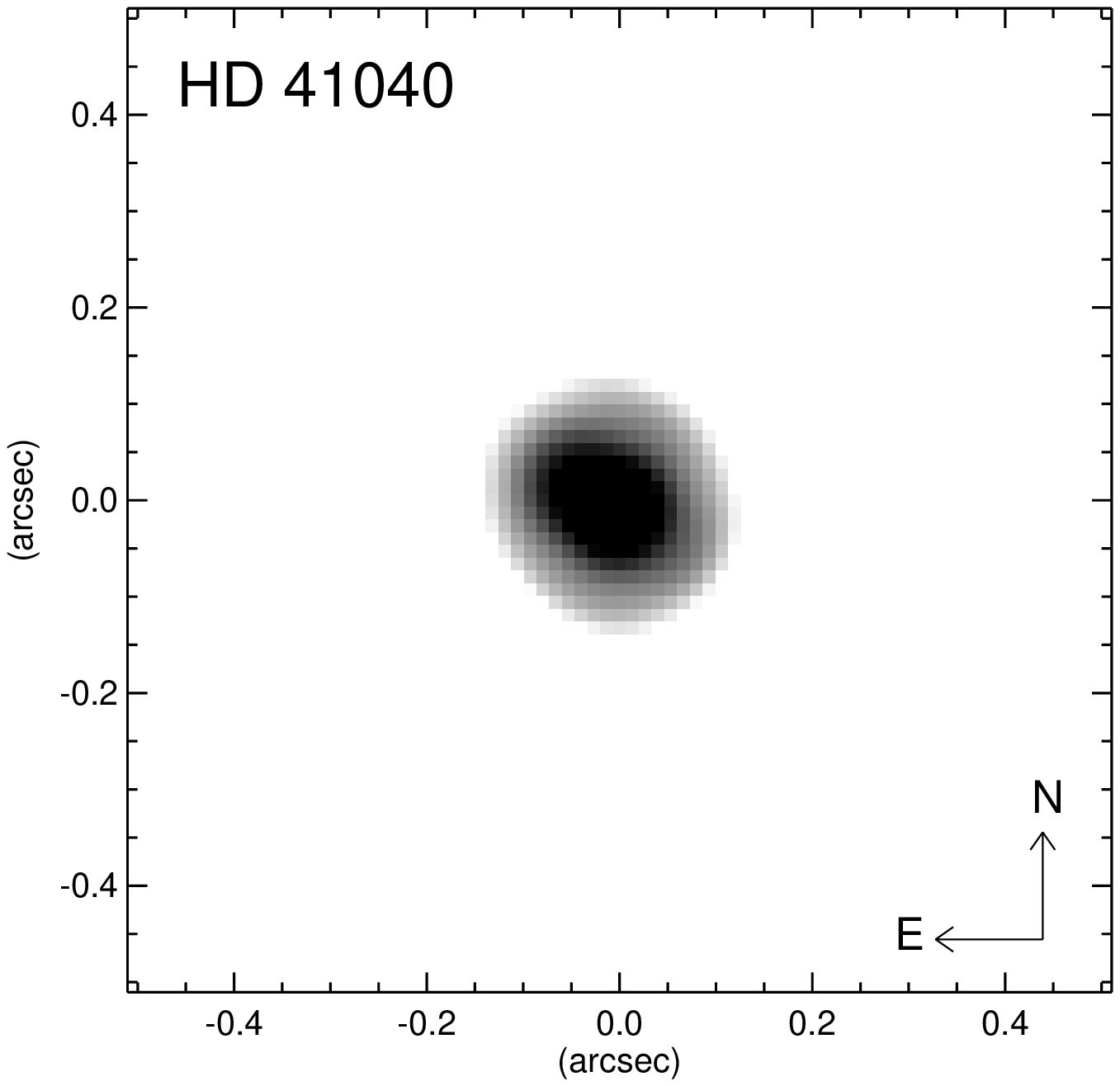}
\includegraphics[width=0.115\textwidth, angle=0]{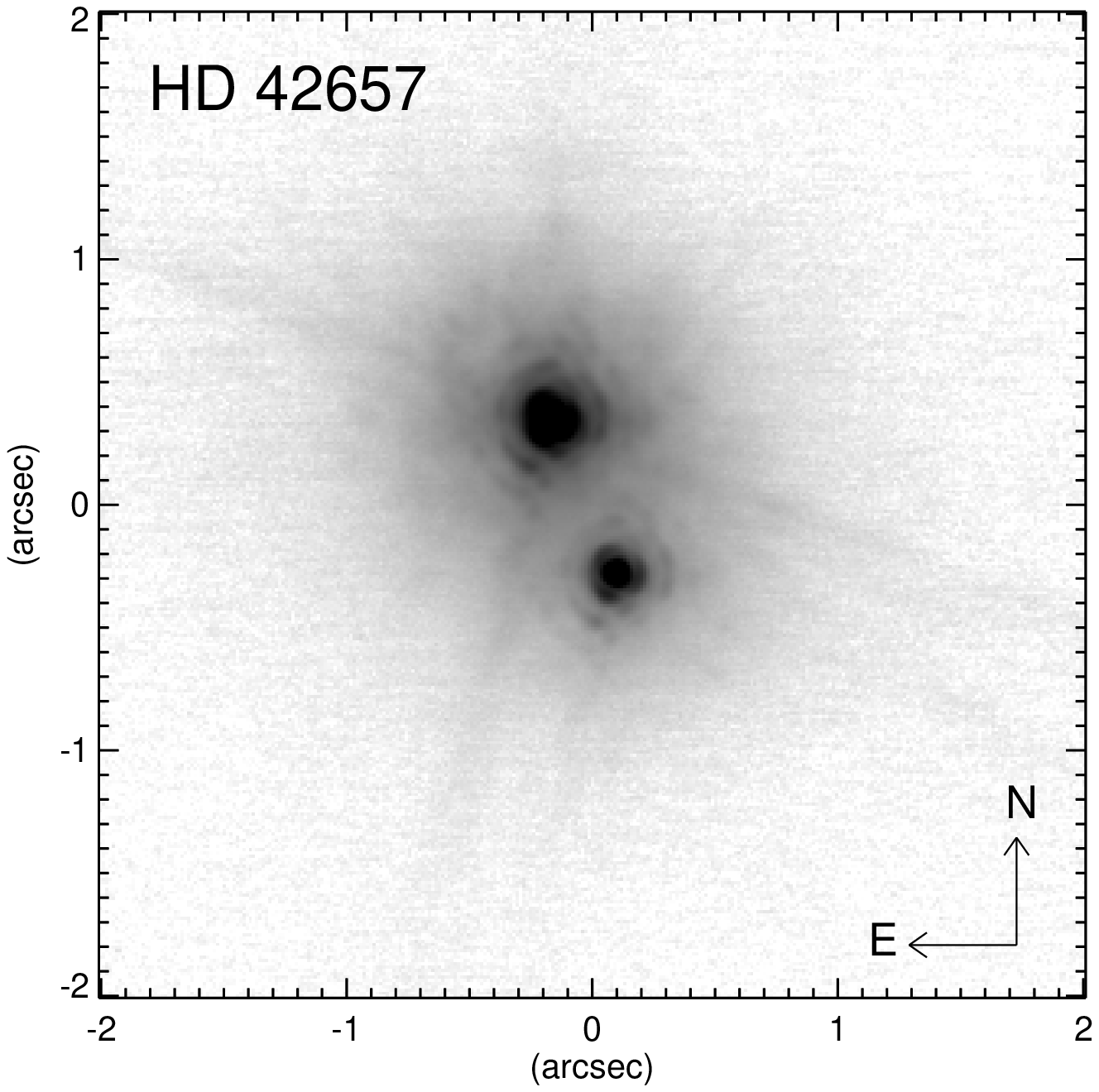}
\includegraphics[width=0.115\textwidth, angle=0]{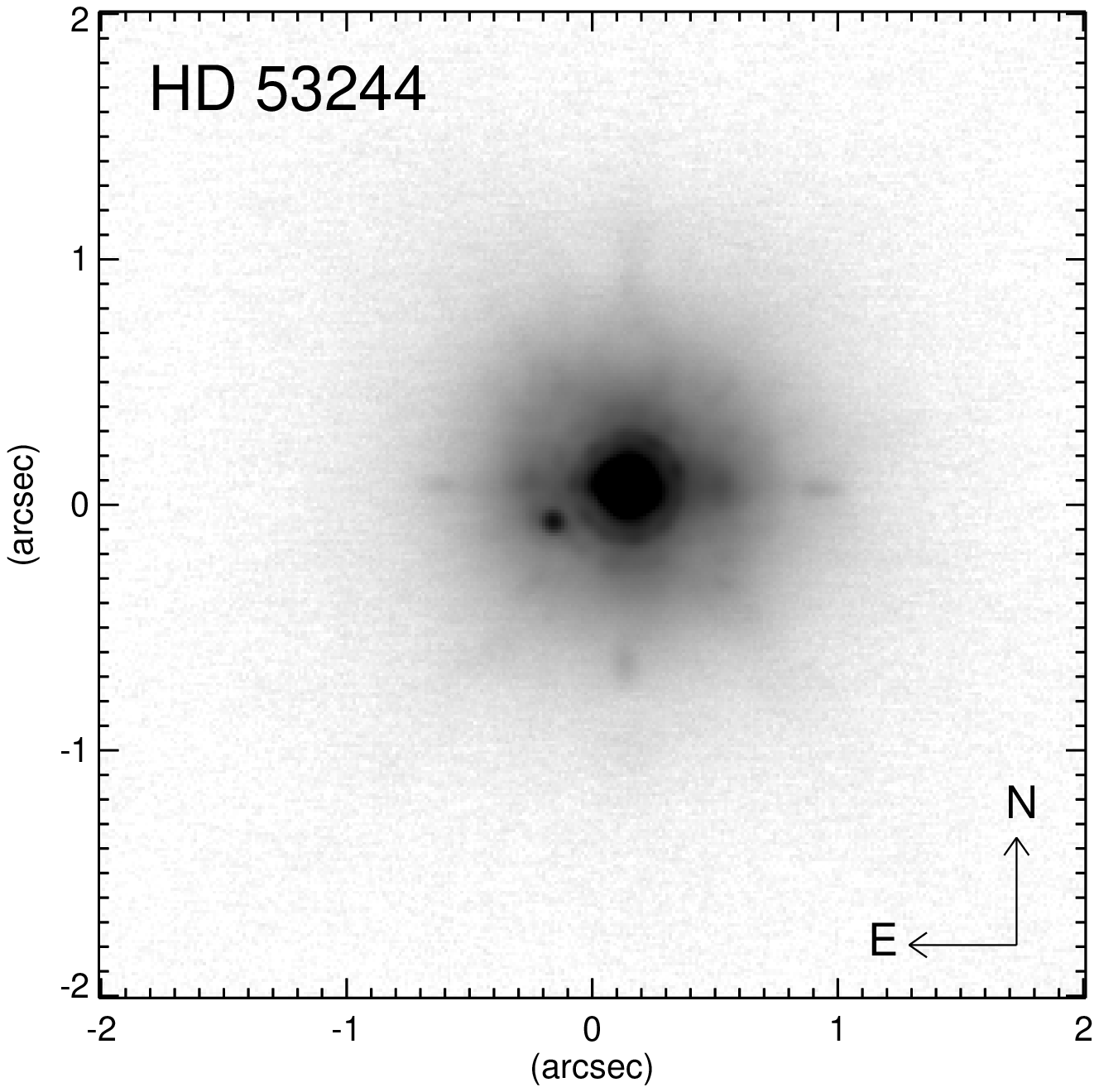}
\includegraphics[width=0.115\textwidth, angle=0]{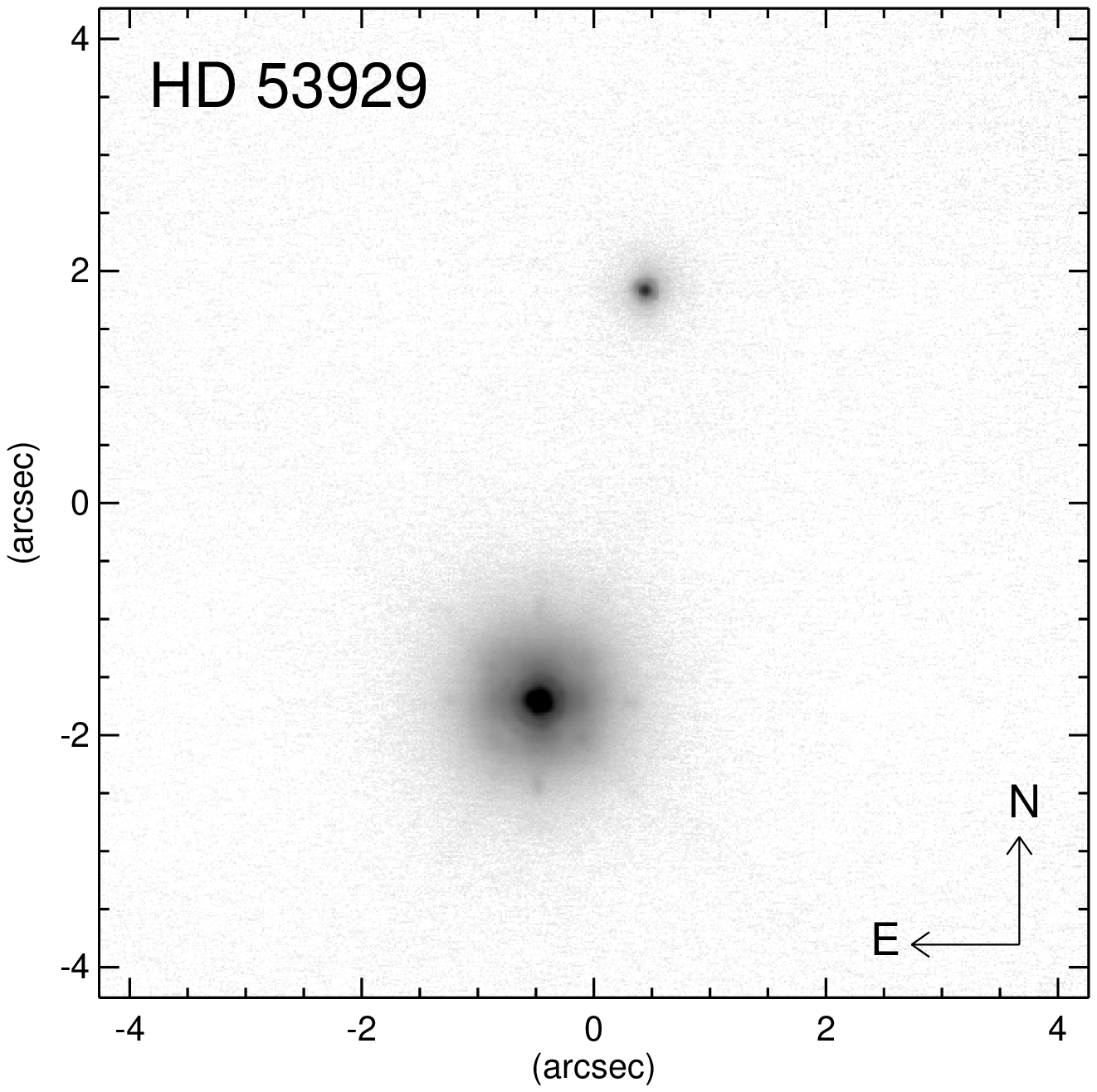}
\includegraphics[width=0.115\textwidth, angle=0]{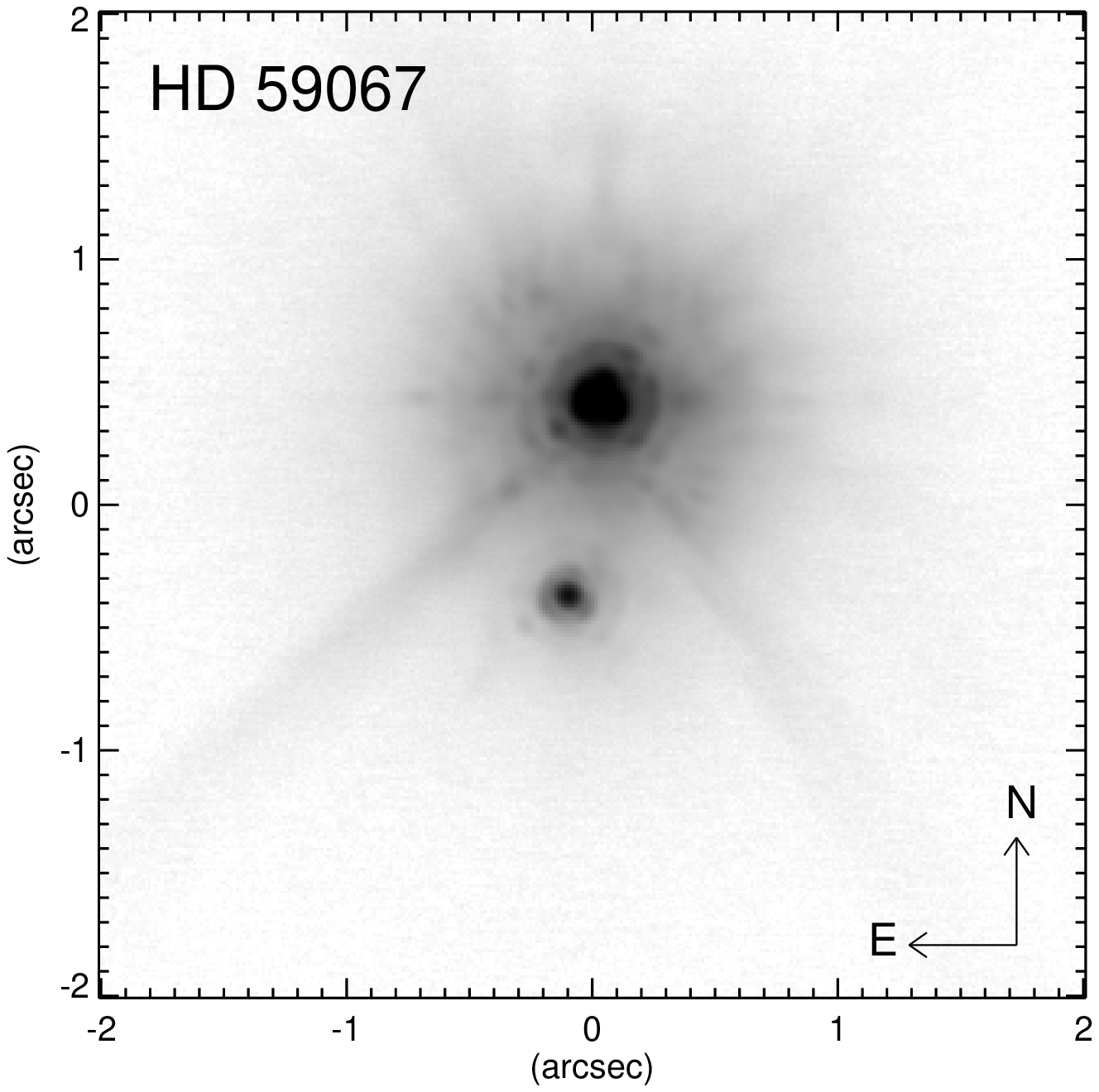}
\includegraphics[width=0.115\textwidth, angle=0]{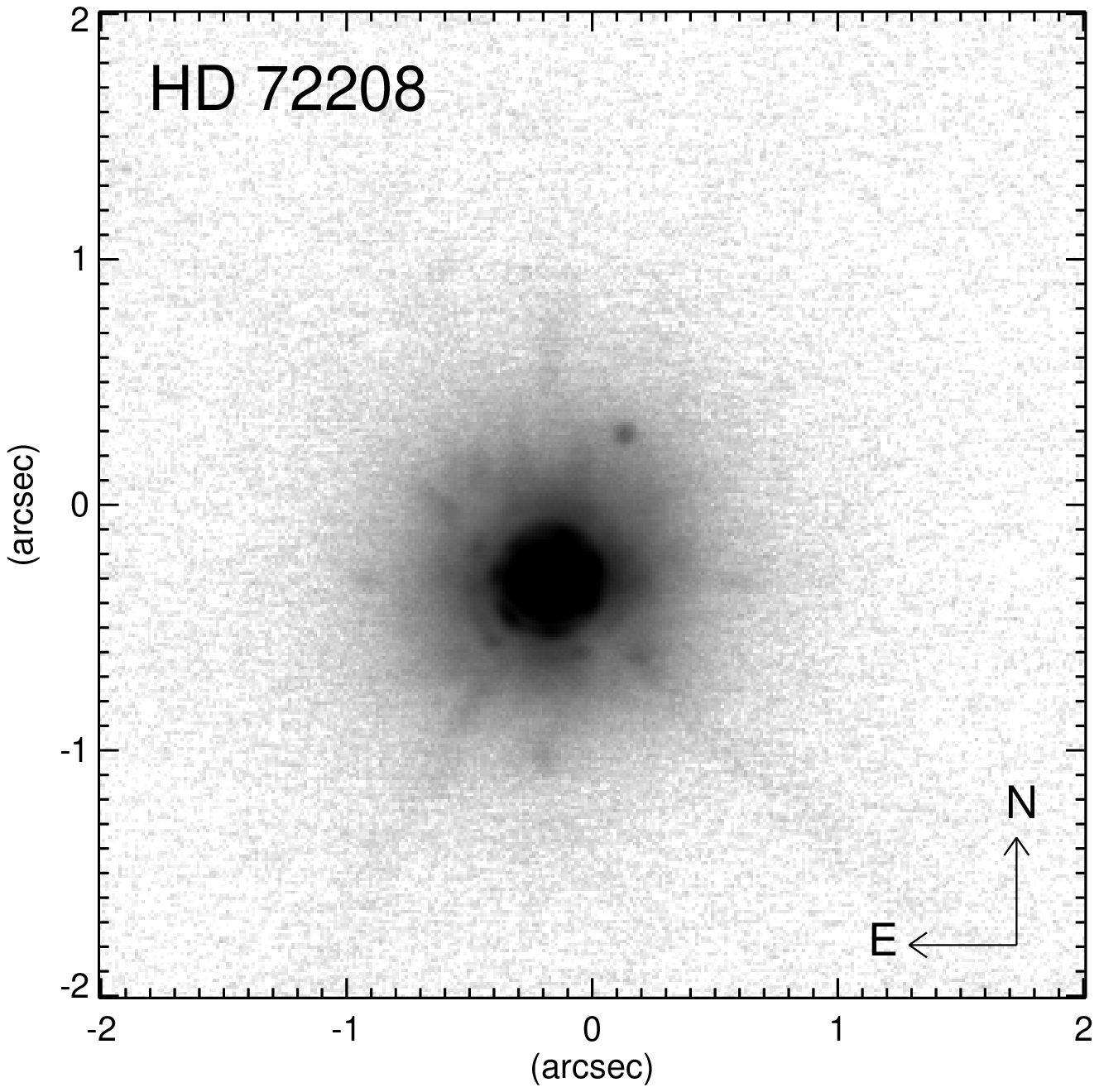}
\includegraphics[width=0.115\textwidth, angle=0]{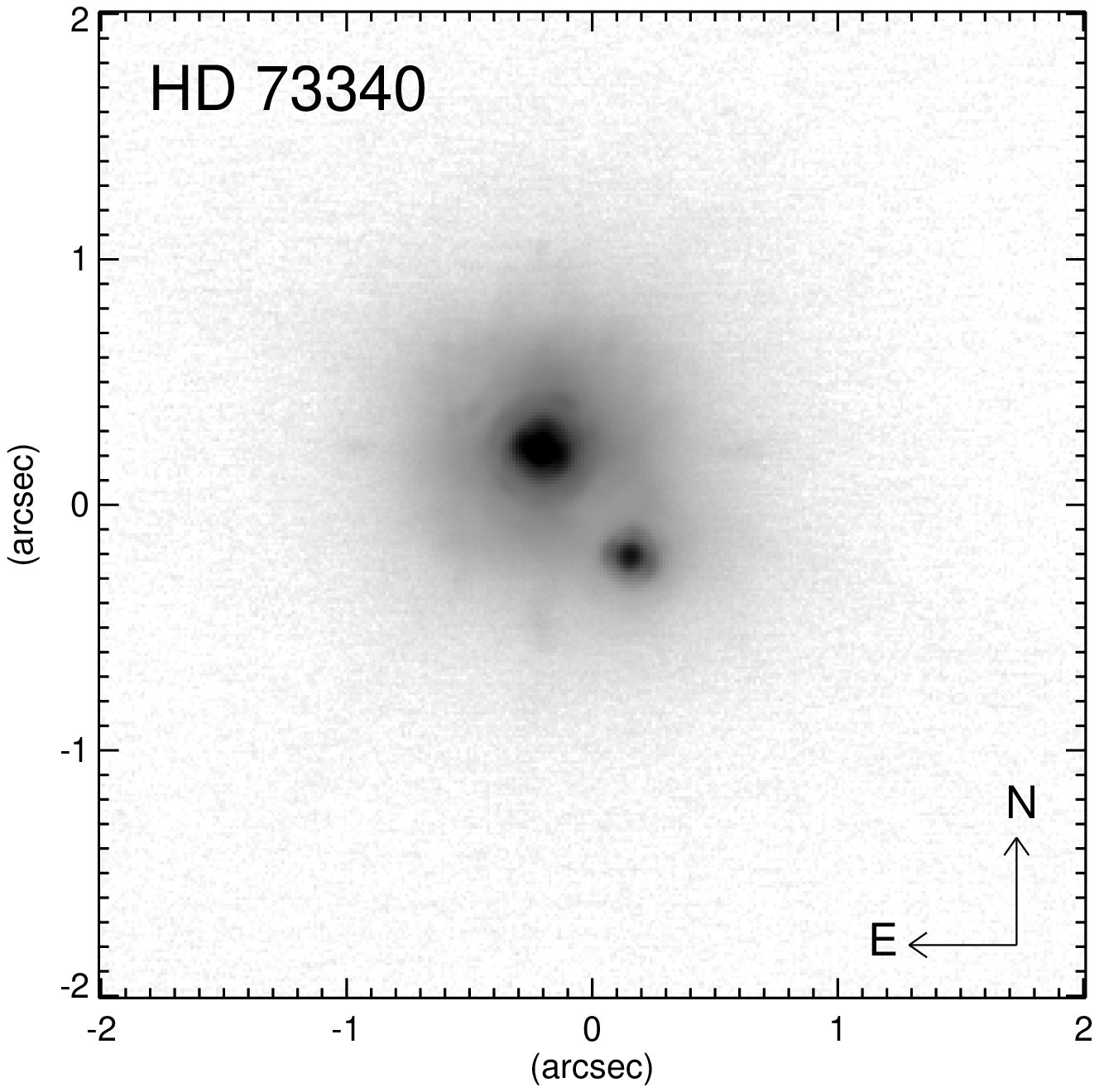}
\includegraphics[width=0.115\textwidth, angle=0]{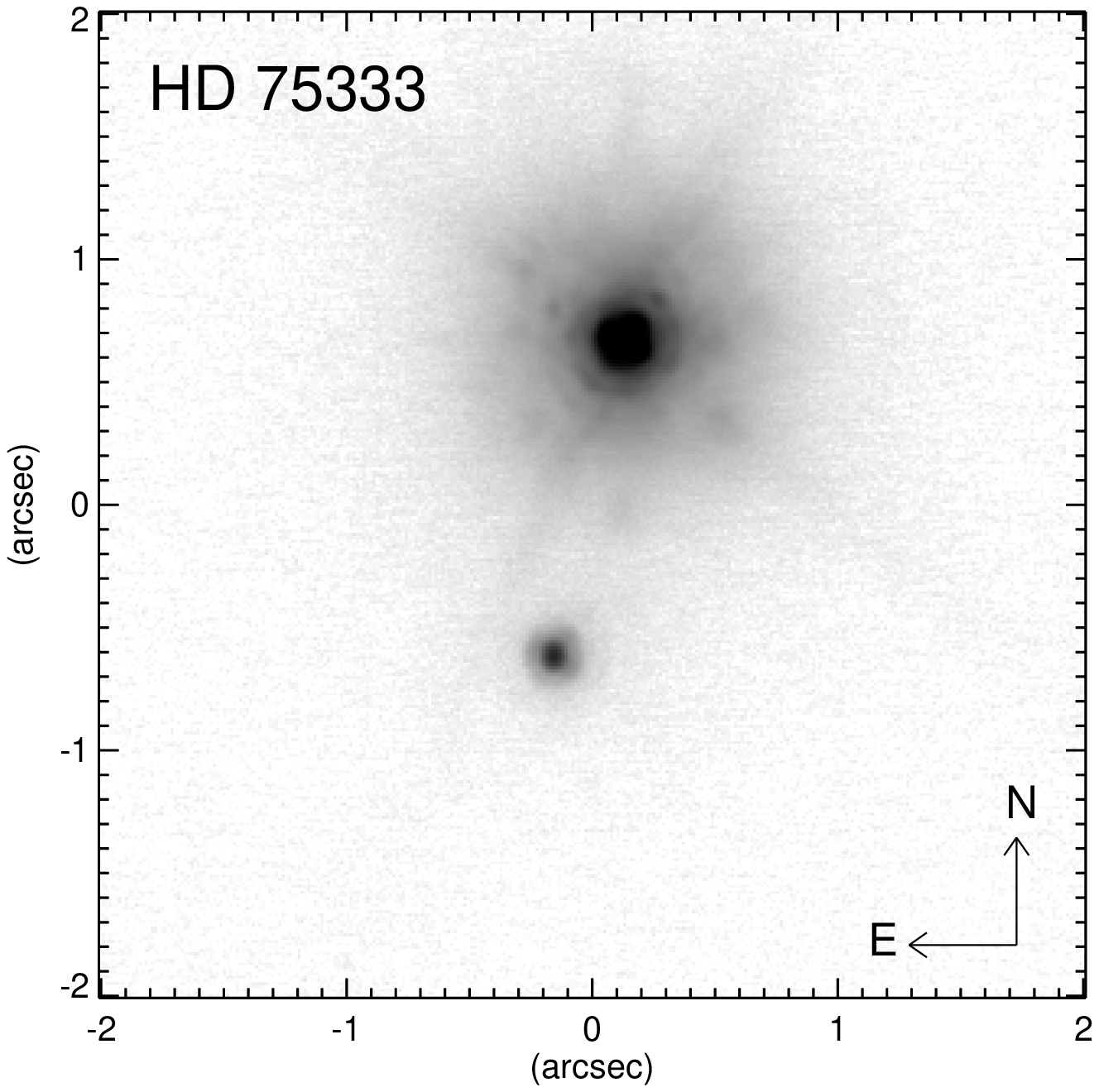}
\includegraphics[width=0.115\textwidth, angle=0]{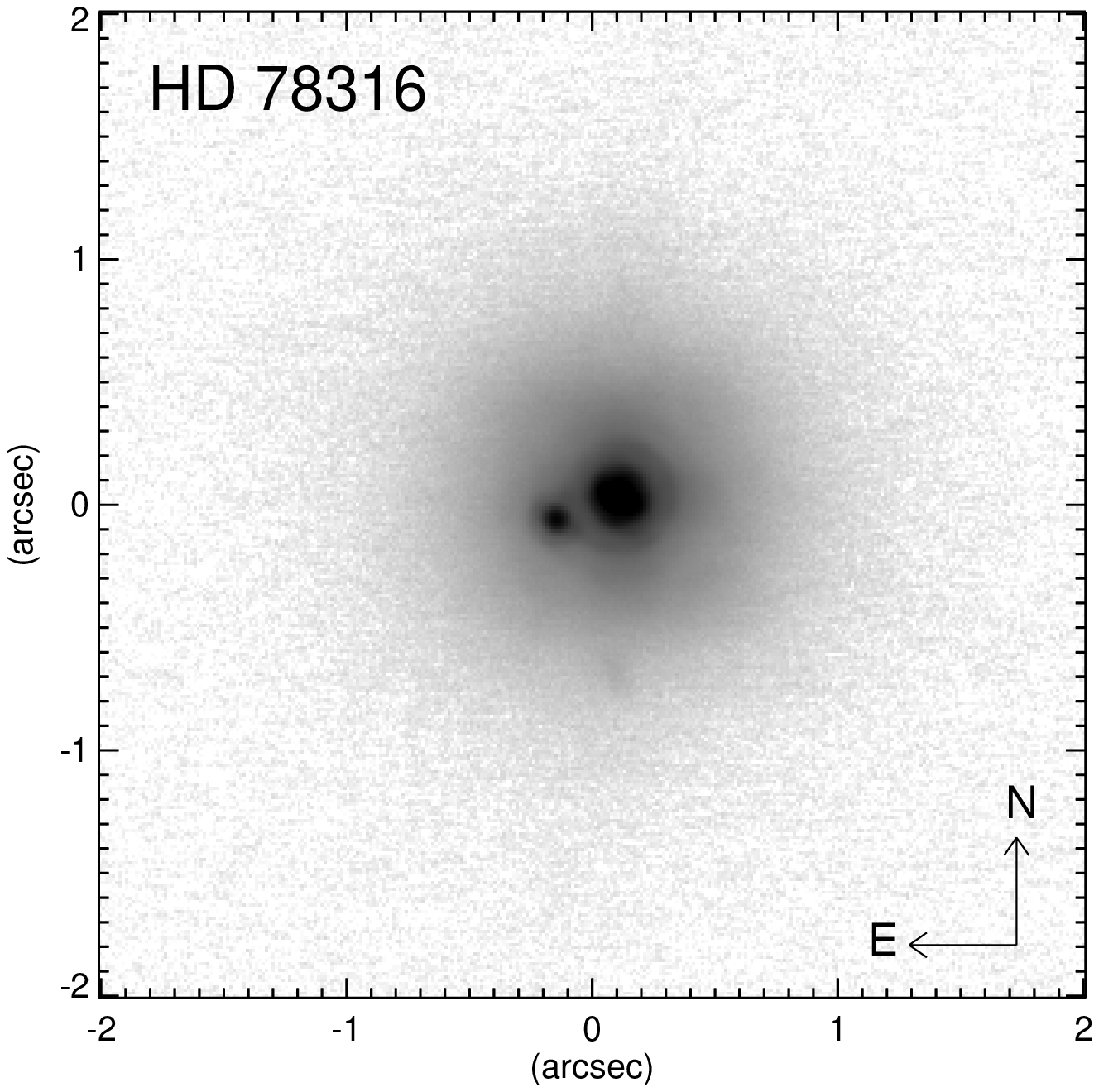}
\includegraphics[width=0.115\textwidth, angle=0]{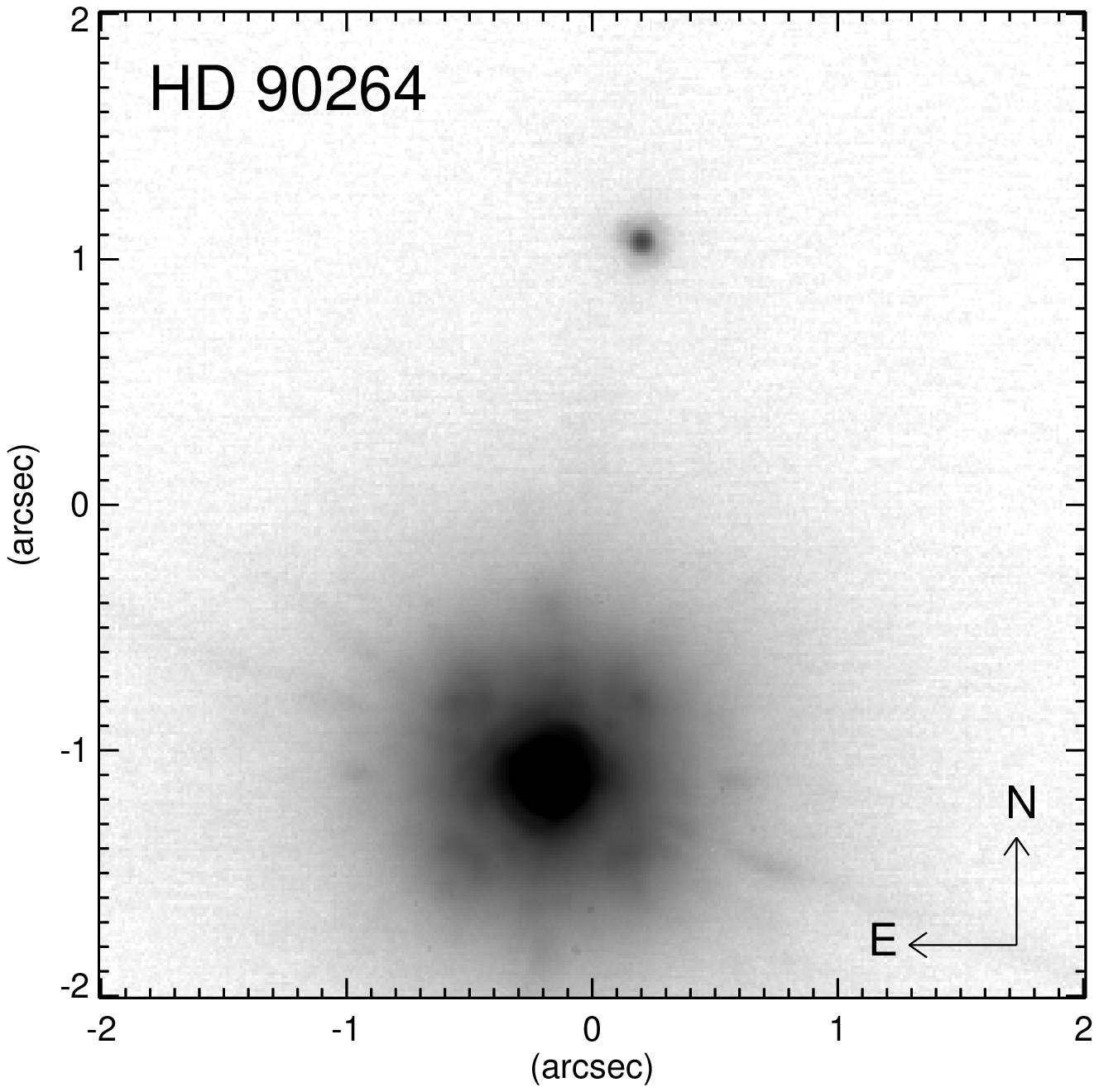}
\includegraphics[width=0.115\textwidth, angle=0]{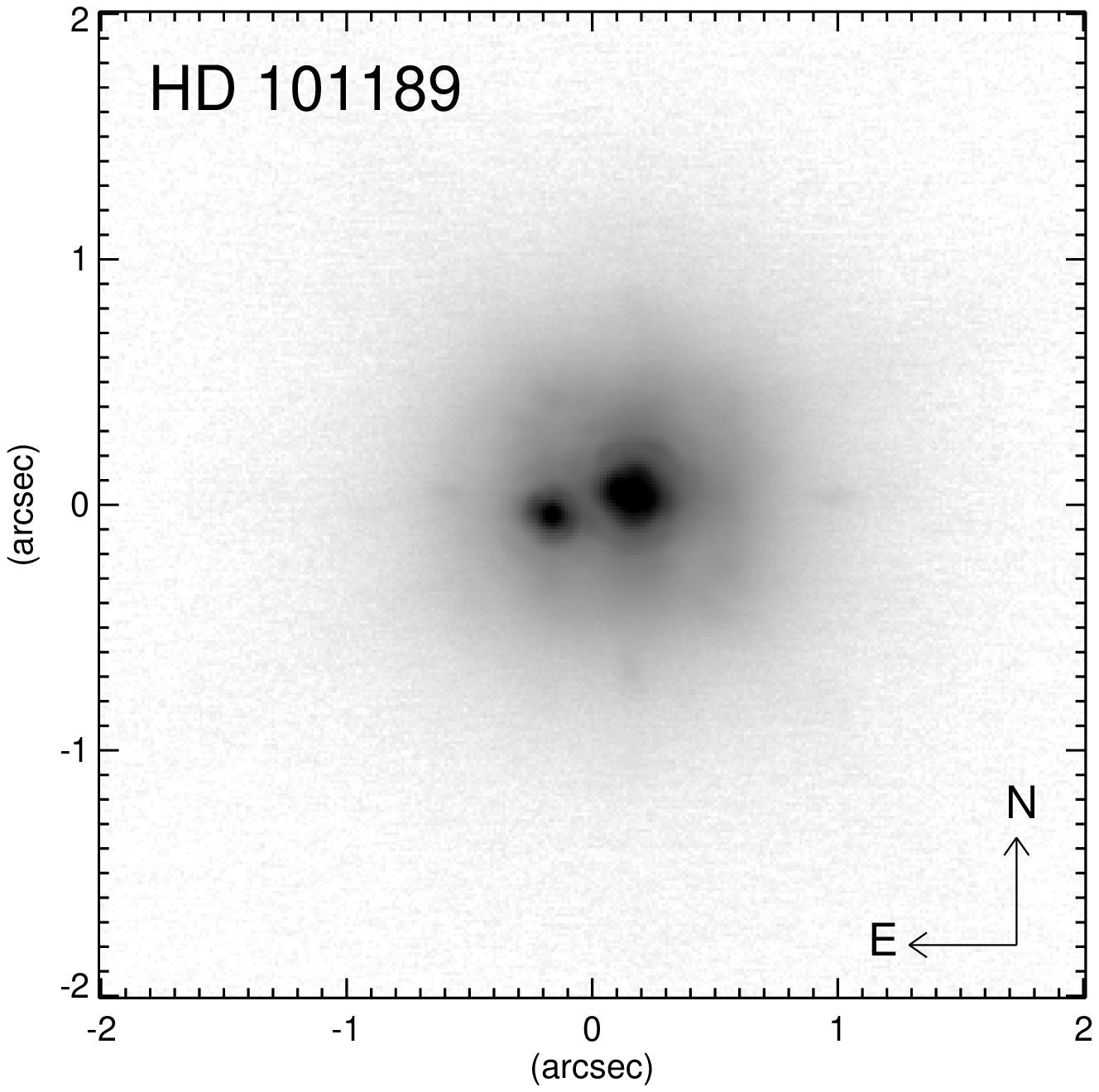}
\includegraphics[width=0.115\textwidth, angle=0]{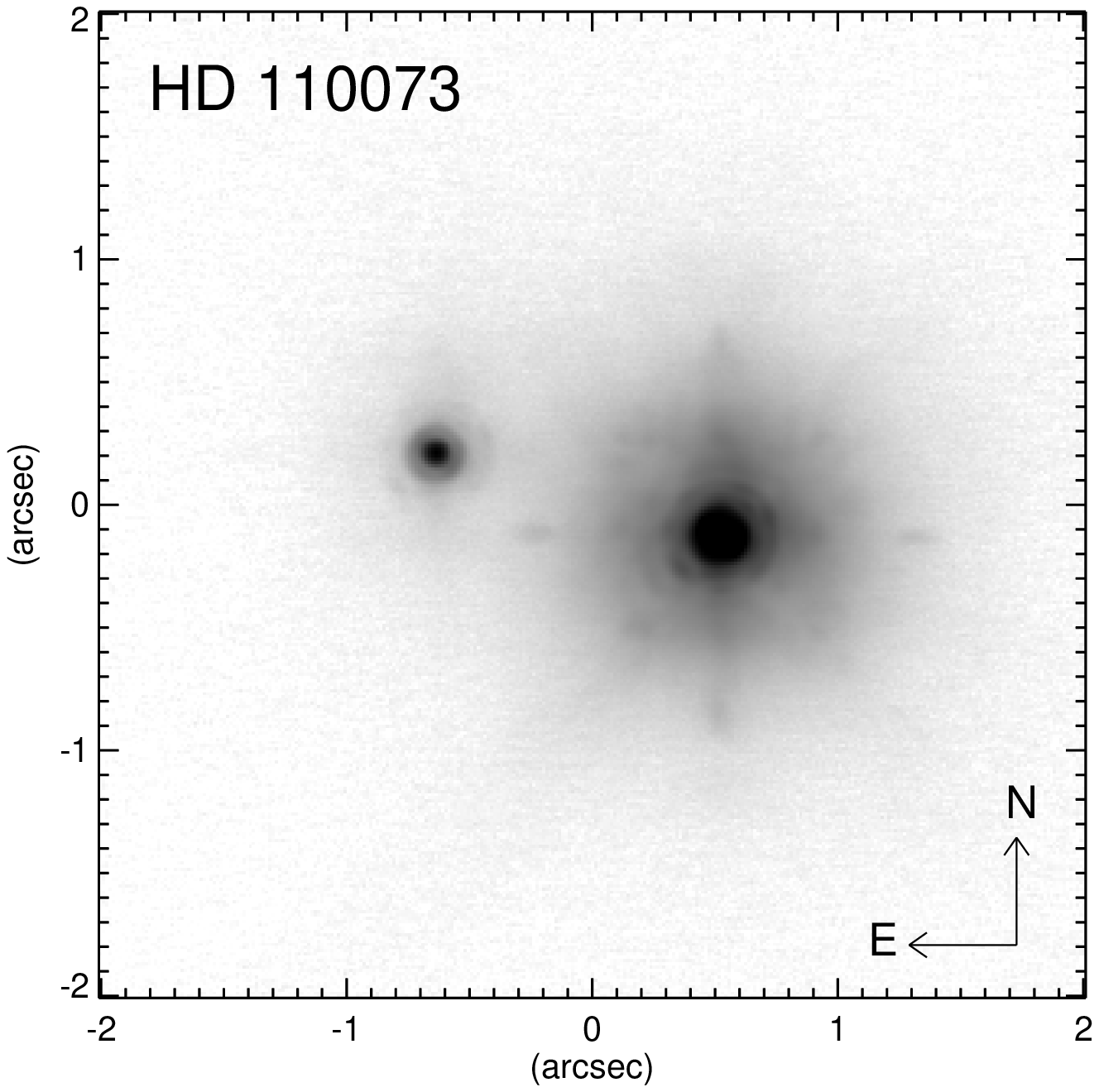}
\includegraphics[width=0.115\textwidth, angle=0]{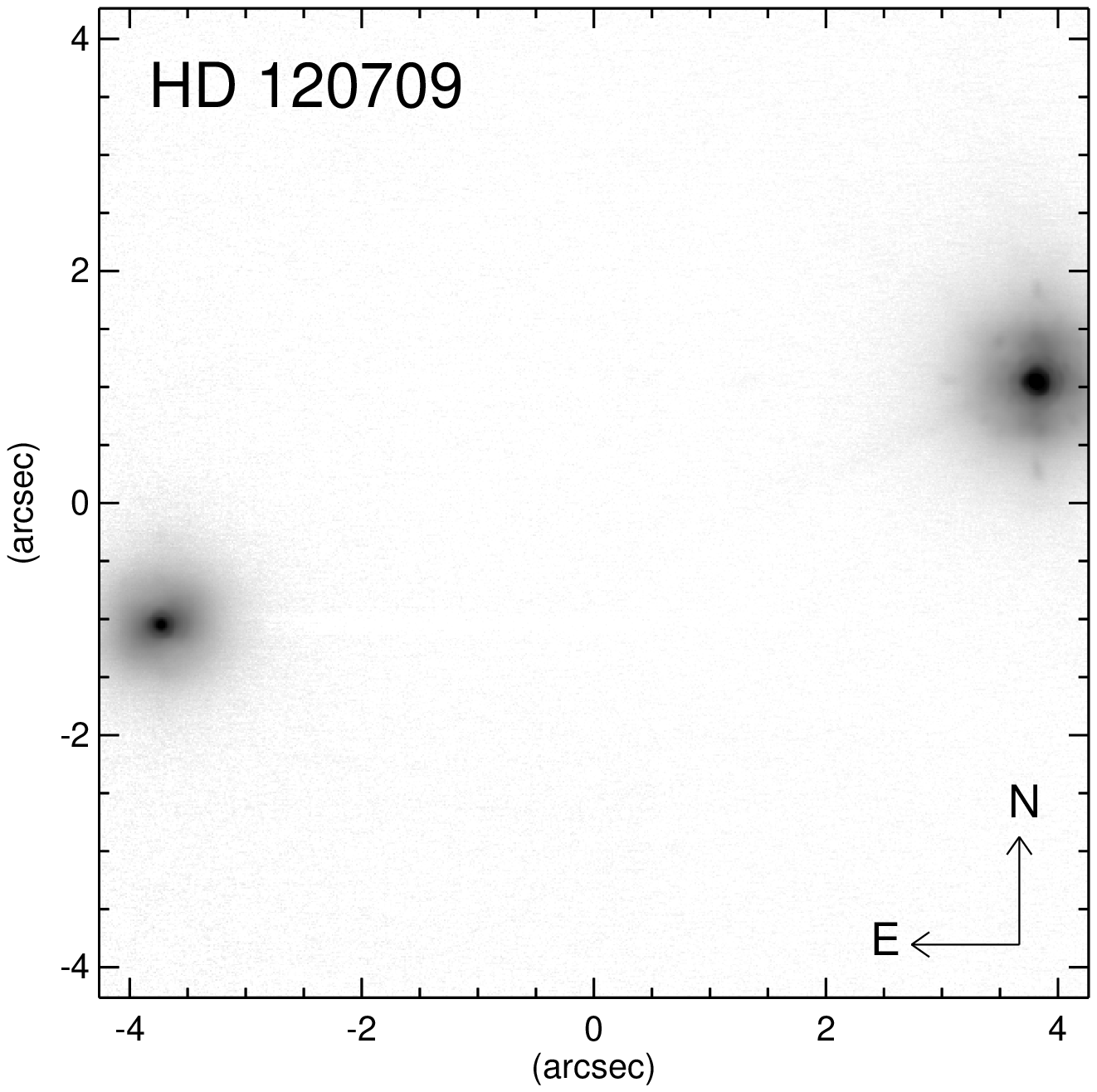}
\includegraphics[width=0.115\textwidth, angle=0]{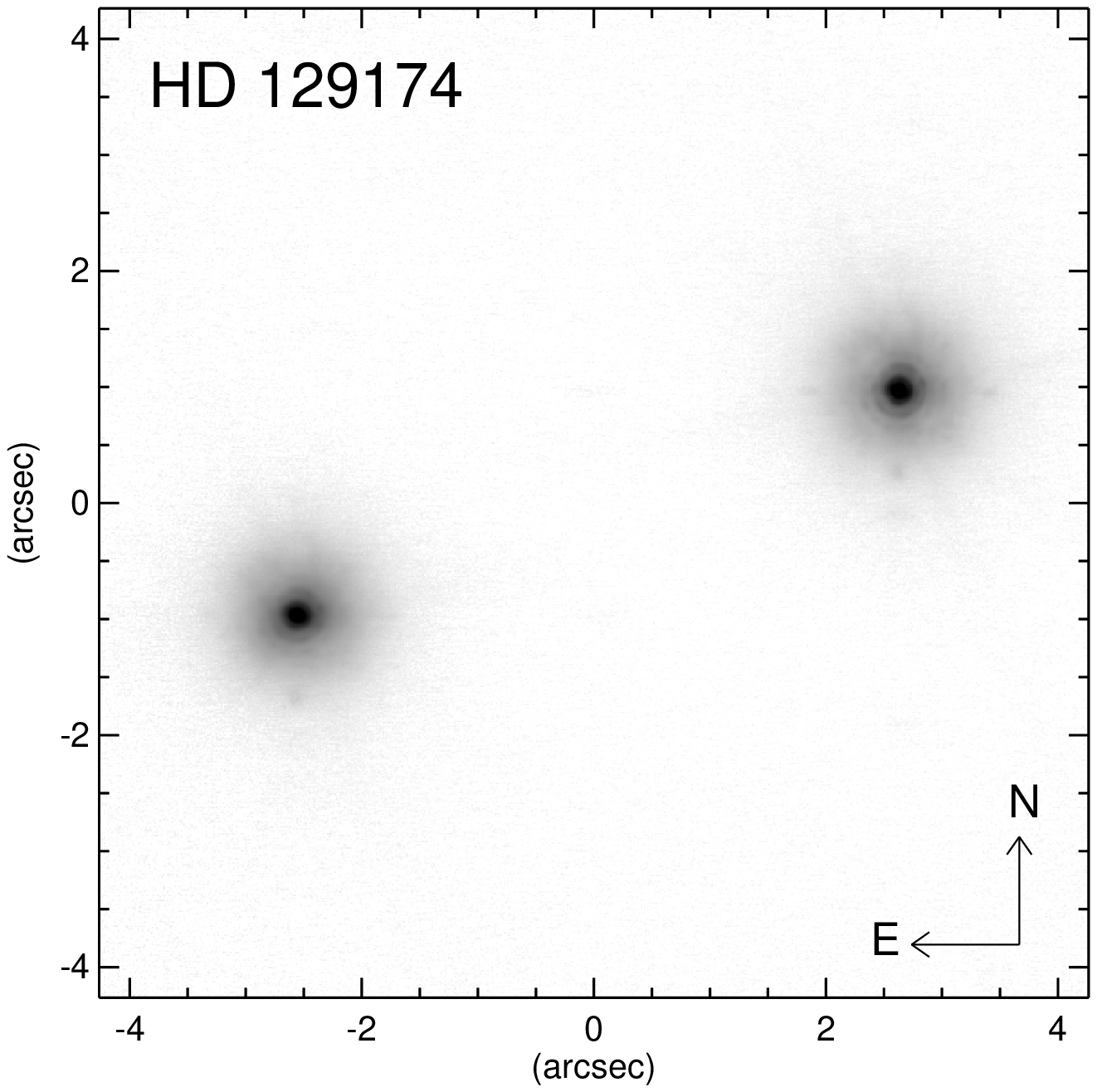}
\includegraphics[width=0.115\textwidth, angle=0]{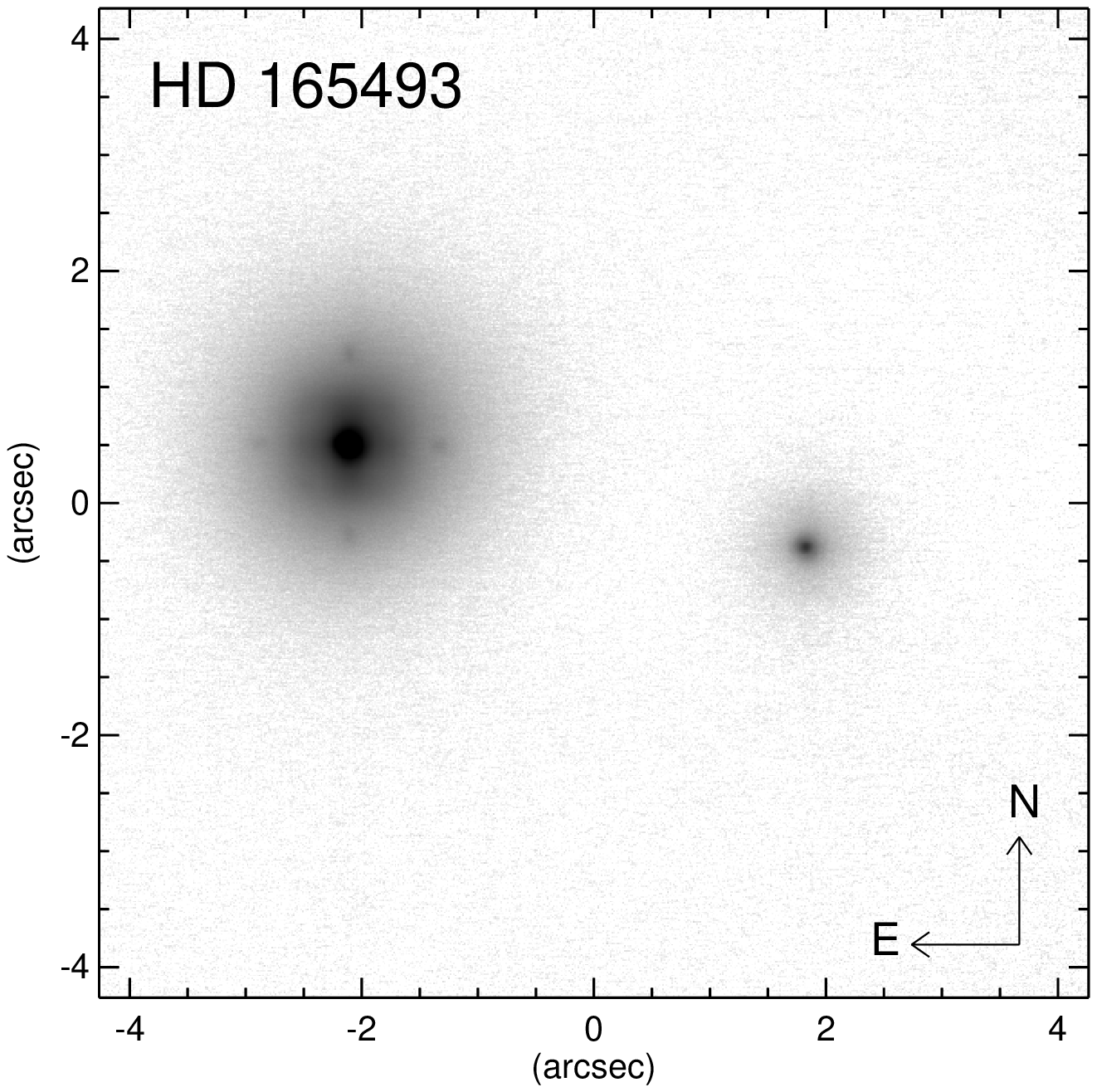}
\includegraphics[width=0.115\textwidth, angle=0]{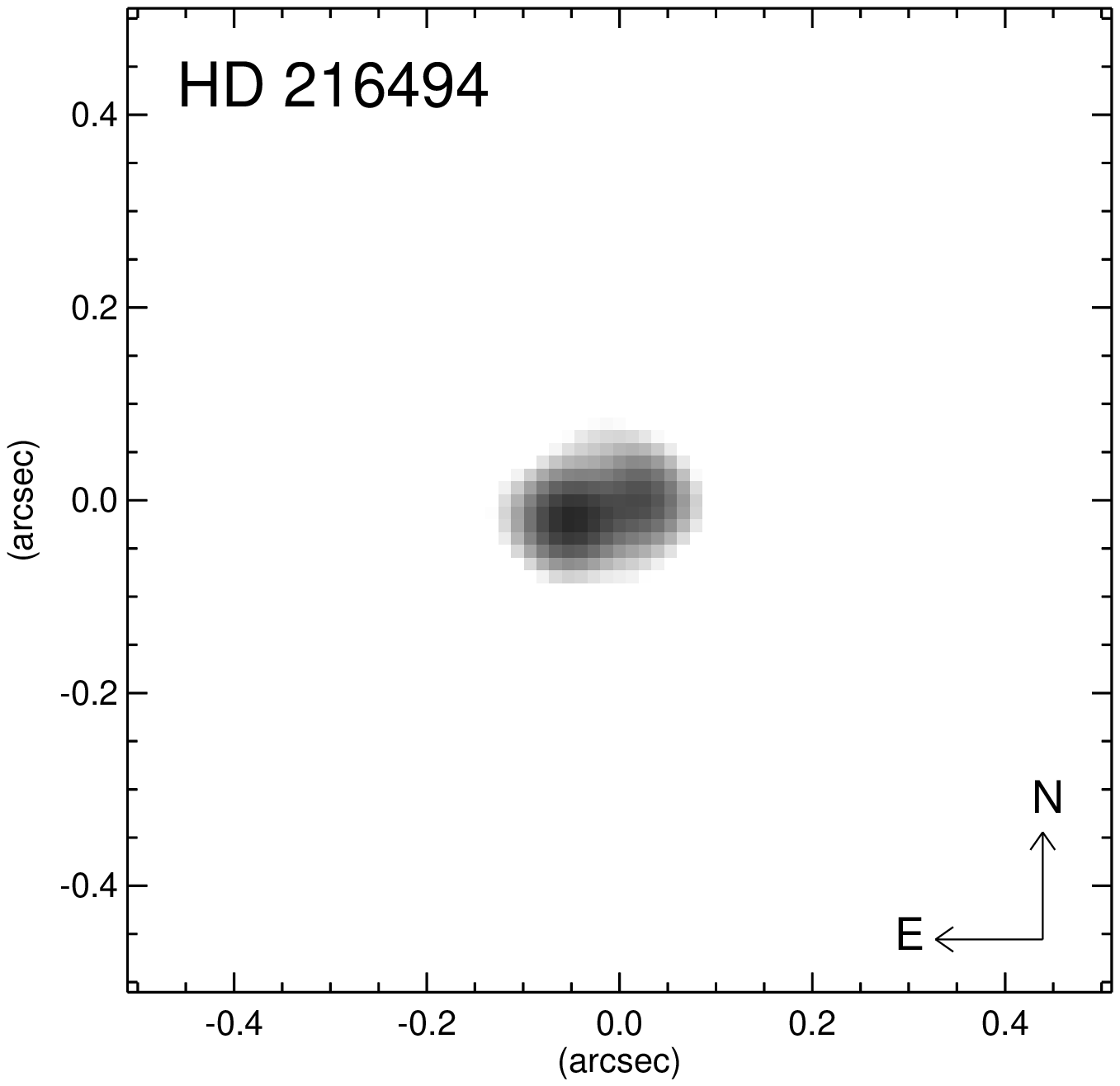}
\includegraphics[width=0.115\textwidth, angle=0]{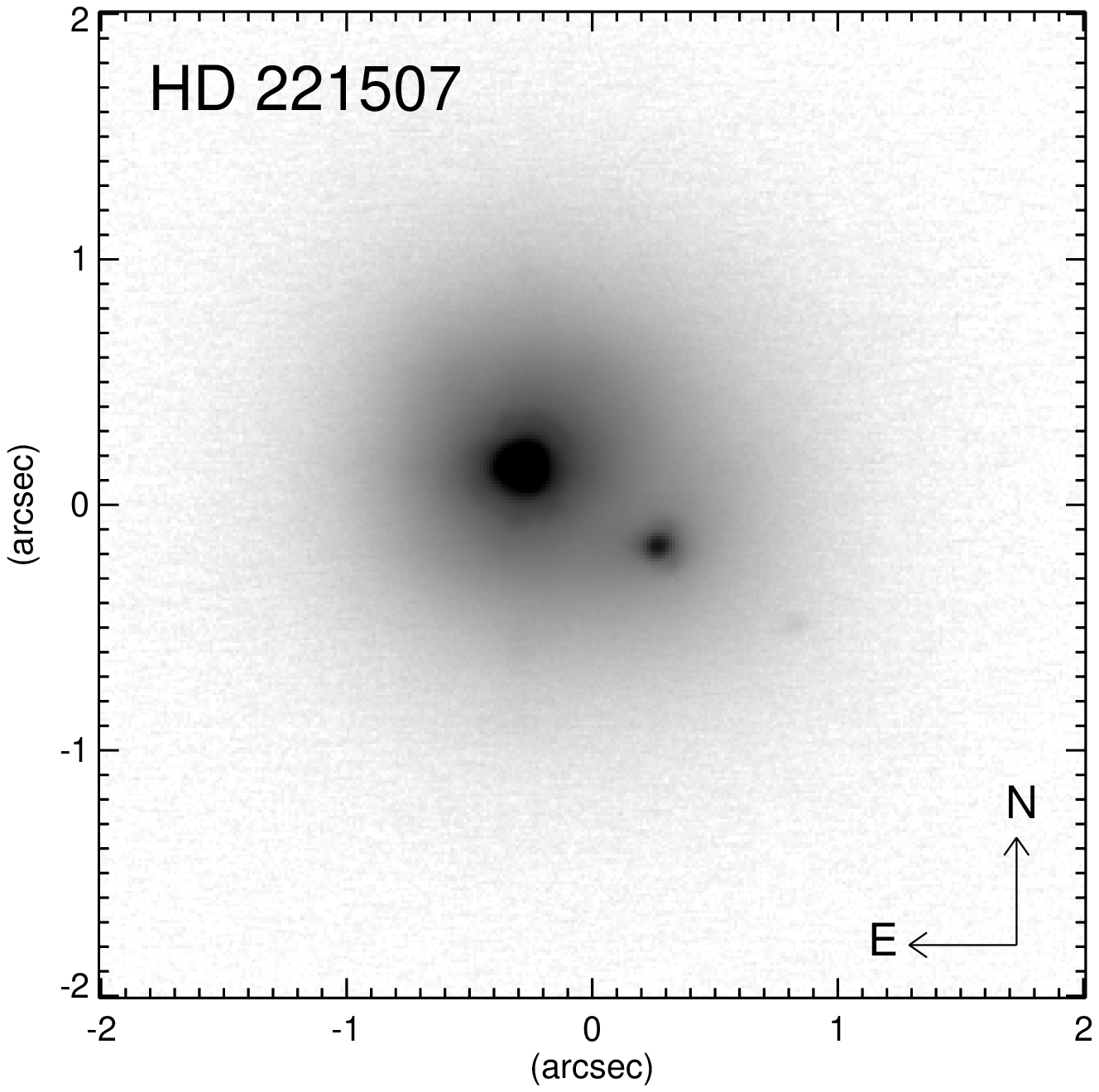}
\includegraphics[width=0.115\textwidth, angle=0]{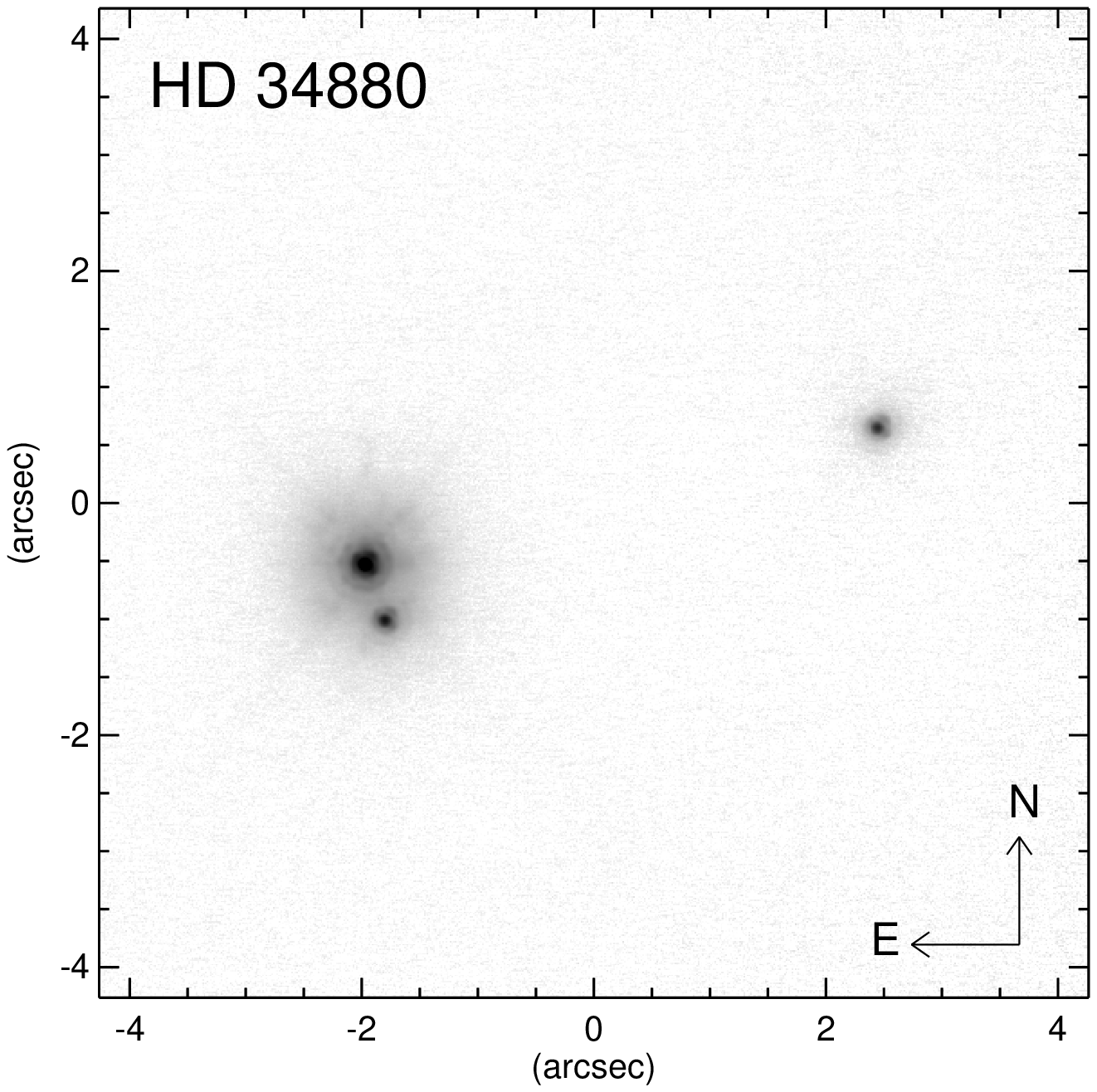}
\includegraphics[width=0.115\textwidth, angle=0]{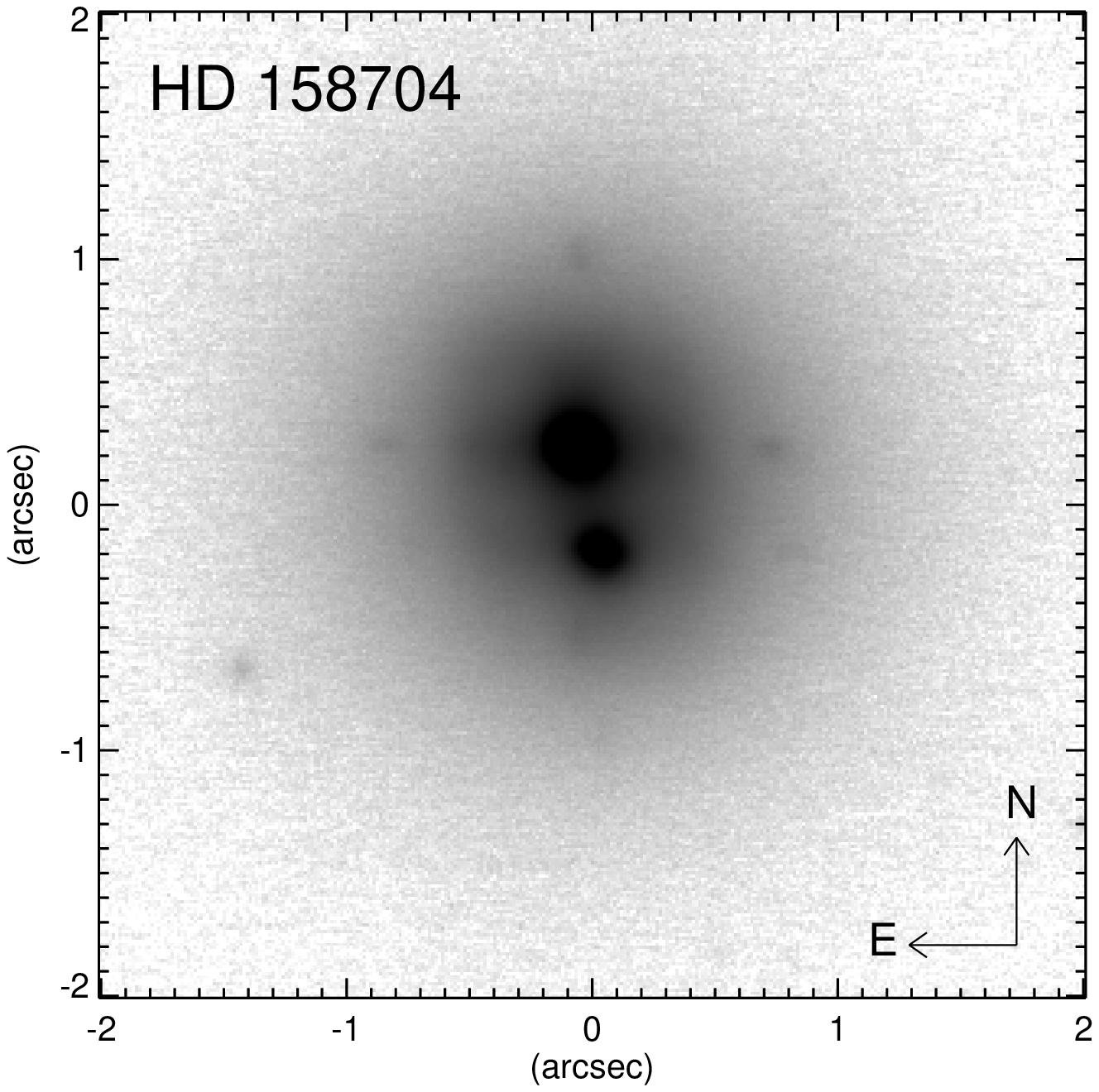}
\includegraphics[width=0.115\textwidth, angle=0]{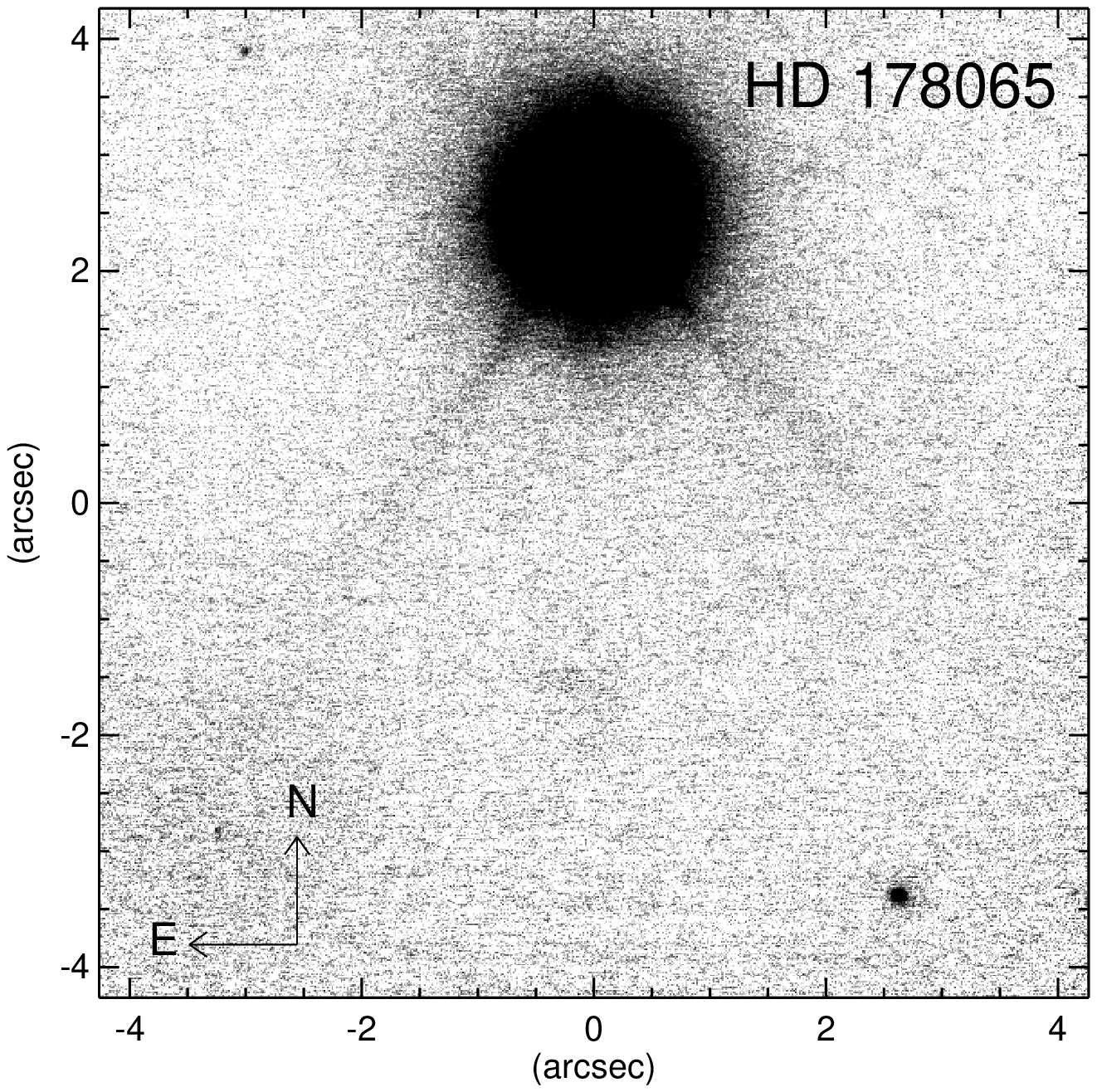}
\includegraphics[width=0.115\textwidth, angle=0]{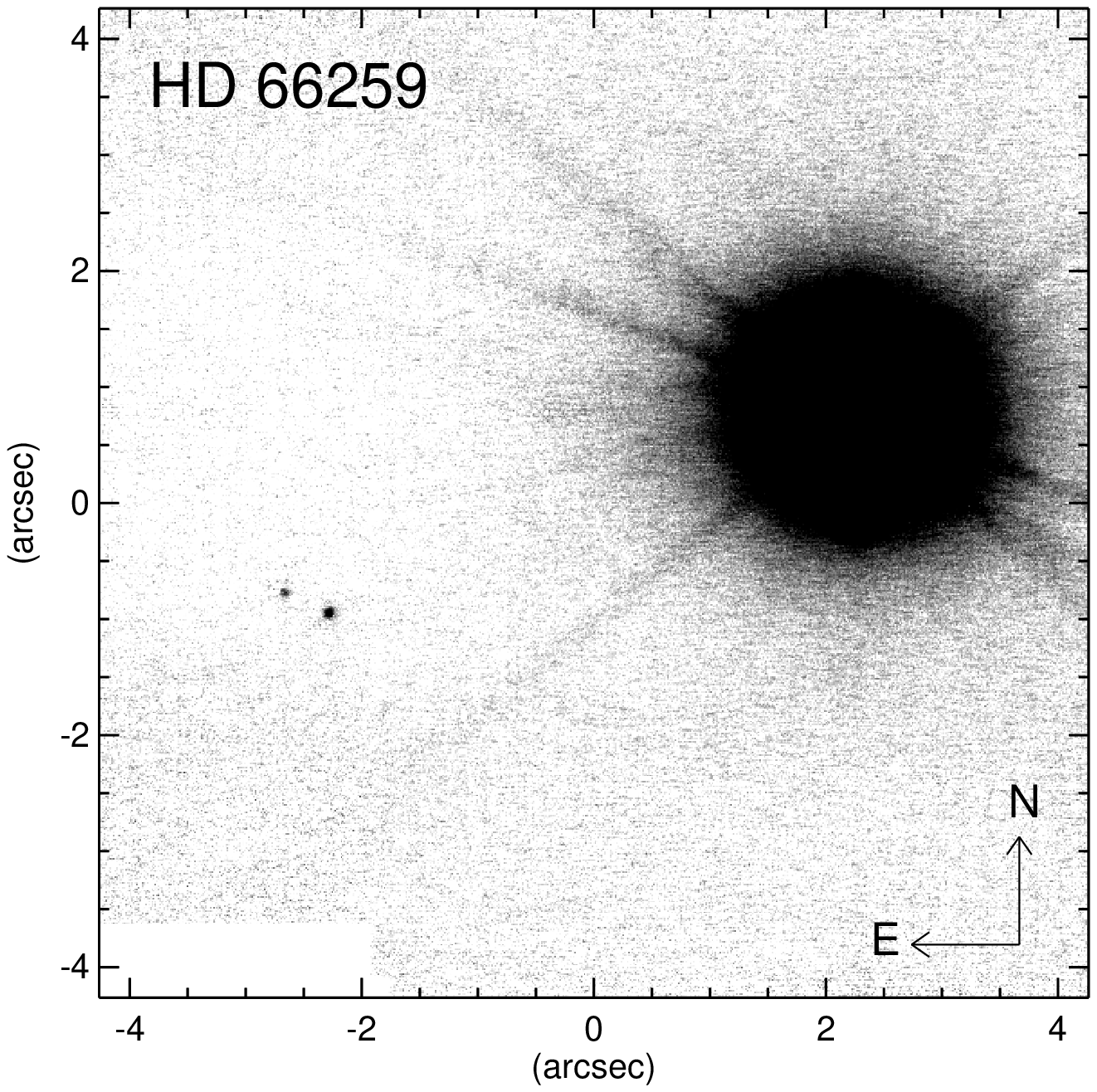}
\includegraphics[width=0.115\textwidth, angle=0]{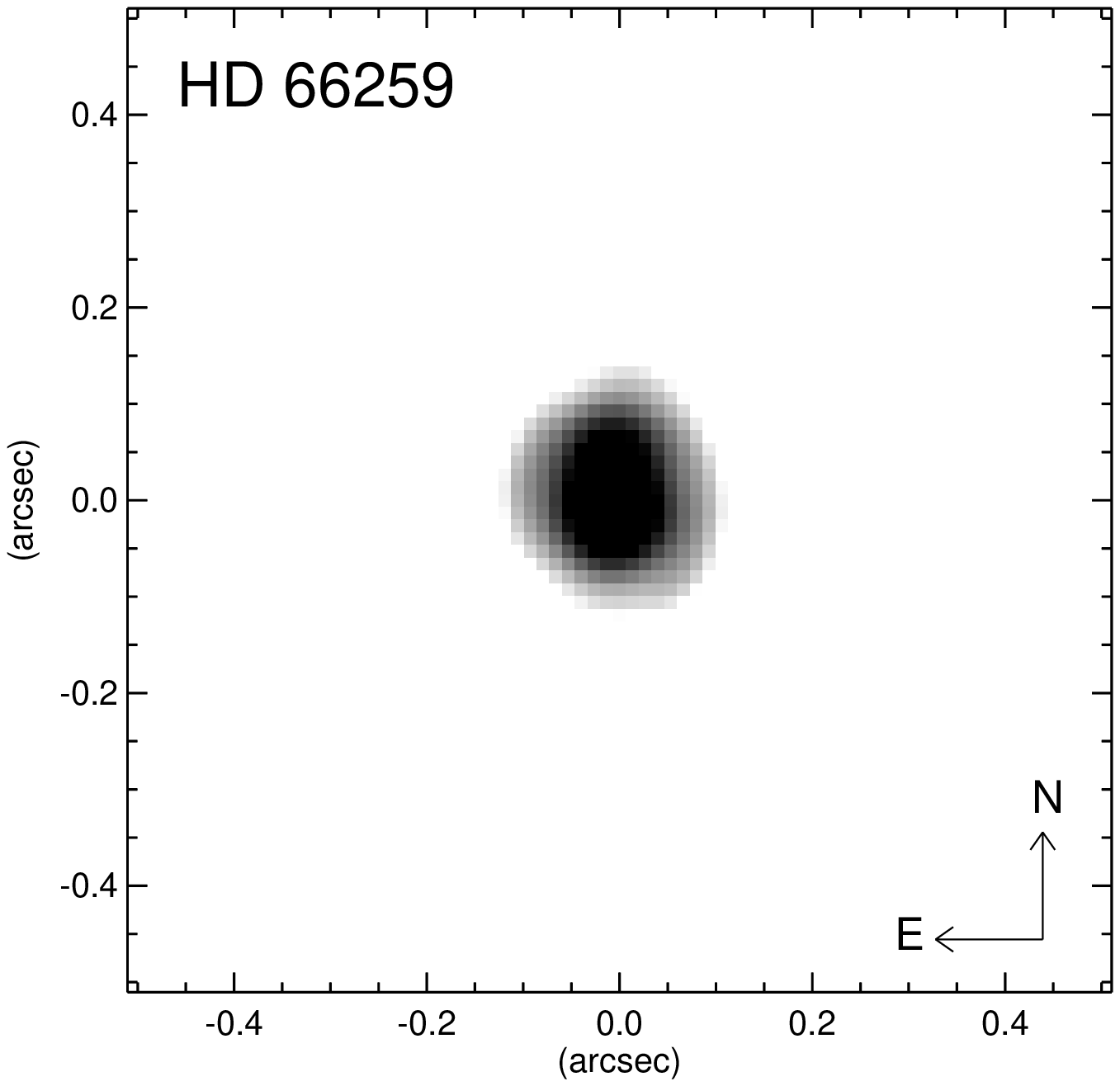}
\caption{
Companion candidates to HgMn stars, detected with NACO by Sch\"oller et al.\ (2010).
-- Credit: Sch\"oller et al., A\&A, 522, A85, 2010, reproduced with permission \copyright ESO.
}
\label{fig:companion_zoo}
\end{figure}

\begin{table}[t]
\centering
\caption{
Multiplicity of different stellar types.
}
\label{tab:binarity}
\begin{tabular}{crlc}
\hline
\hline
\multicolumn{1}{c}{Type} &
\multicolumn{1}{c}{Percentage} &
\multicolumn{1}{c}{Reference} &
\multicolumn{1}{c}{SB} \\
\hline
Normal A & $\sim35$\% & Kouwenhoven et al.\ 2005 &  \\
Normal B & $\sim30$\% & Kouwenhoven et al.\ 2005 &  \\
Magnetic Ap & 43\% & Carrier et al.\ 2002 & Very few SB2 \\
Magnetic Bp & $\sim20$\% & Renson \& Manfroid 2009 & Very few SB2  \\
HgMn & $>90$\% & Sch\"oller et al.\ 2010 & 2/3  \\
Am & $>90$\% & Renson \& Manfroid 2009 & $>90$\%  \\
roAp & 24\% & Sch\"oller et al.\ 2012 & 2 out of $\sim45$ \\
\hline
\end{tabular}
\end{table}

Sch\"oller et al.\ (2010) studied the 
multiplicity of late-type B stars with HgMn peculiarity.
From observations of 57 HgMn stars obtained at the VLT with the NACO instrument
in Ks  with the S13 camera, 
they found 
34 companion candidates in 25 binaries, three triples, and one quadruple (see Fig.~\ref{fig:companion_zoo}).
Nine companion candidates were found for the first time,
five objects are very likely chance projections.
Only five stars in the total sample show no indication of multiplicity,
taking into account that 44 systems are confirmed or suspected spectroscopic binaries.
On the other hand, in a study of rapidly oscillating Ap (roAp) stars,
Sch\"oller et al.\ (2012) found
a total of six high probability companion candidates in a survey of
28 roAp stars.
roAp stars pulsate in high-overtone, low-degree, nonradial $p$-modes, with periods in the
range from 5.6 to 21\,min and typical amplitudes of a few millimagnitudes
(e.g.\ Kurtz et al.\ 1982).
They are ideal targets for asteroseismology.
The intriguing question is if and how multiplicity can shape the appearance of chemically peculiar stars.
An overview about the prevalence of binaries in different classes of CP stars and normal stars can be
found in Table~\ref{tab:binarity}.

\begin{figure}[t]
\centering
\includegraphics[width=0.32\textwidth]{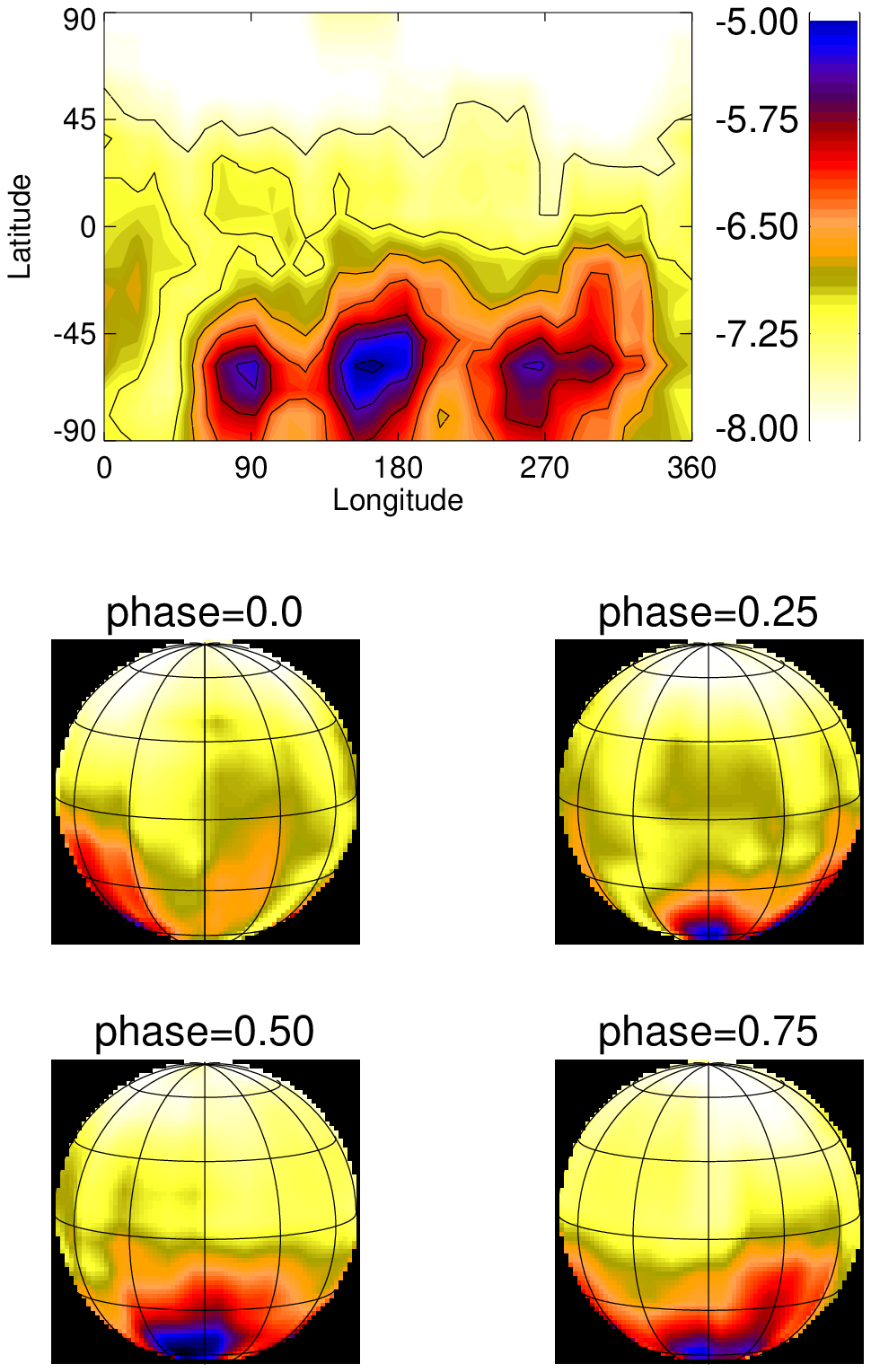}
\includegraphics[width=0.32\textwidth]{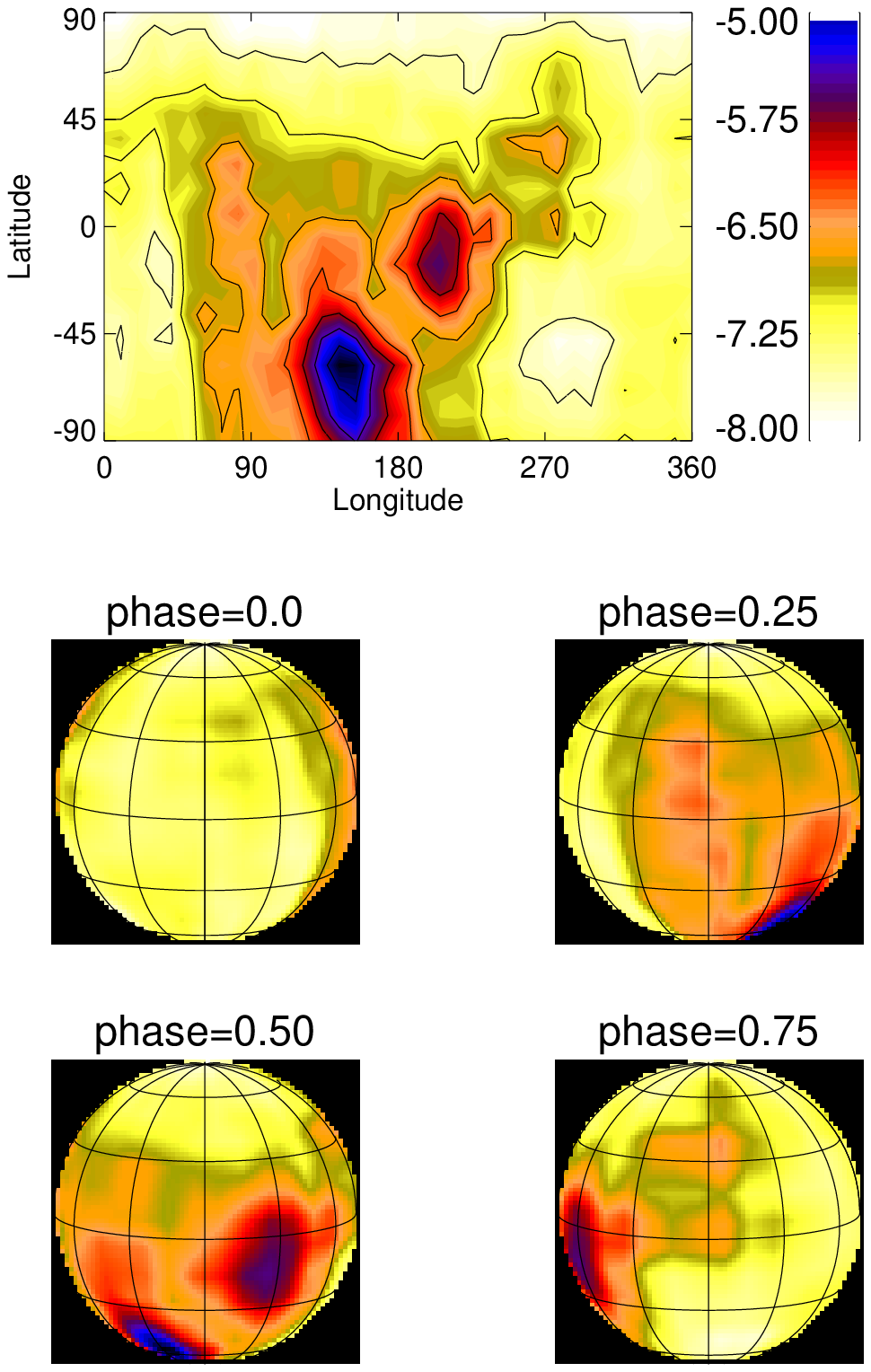}
\includegraphics[width=0.32\textwidth]{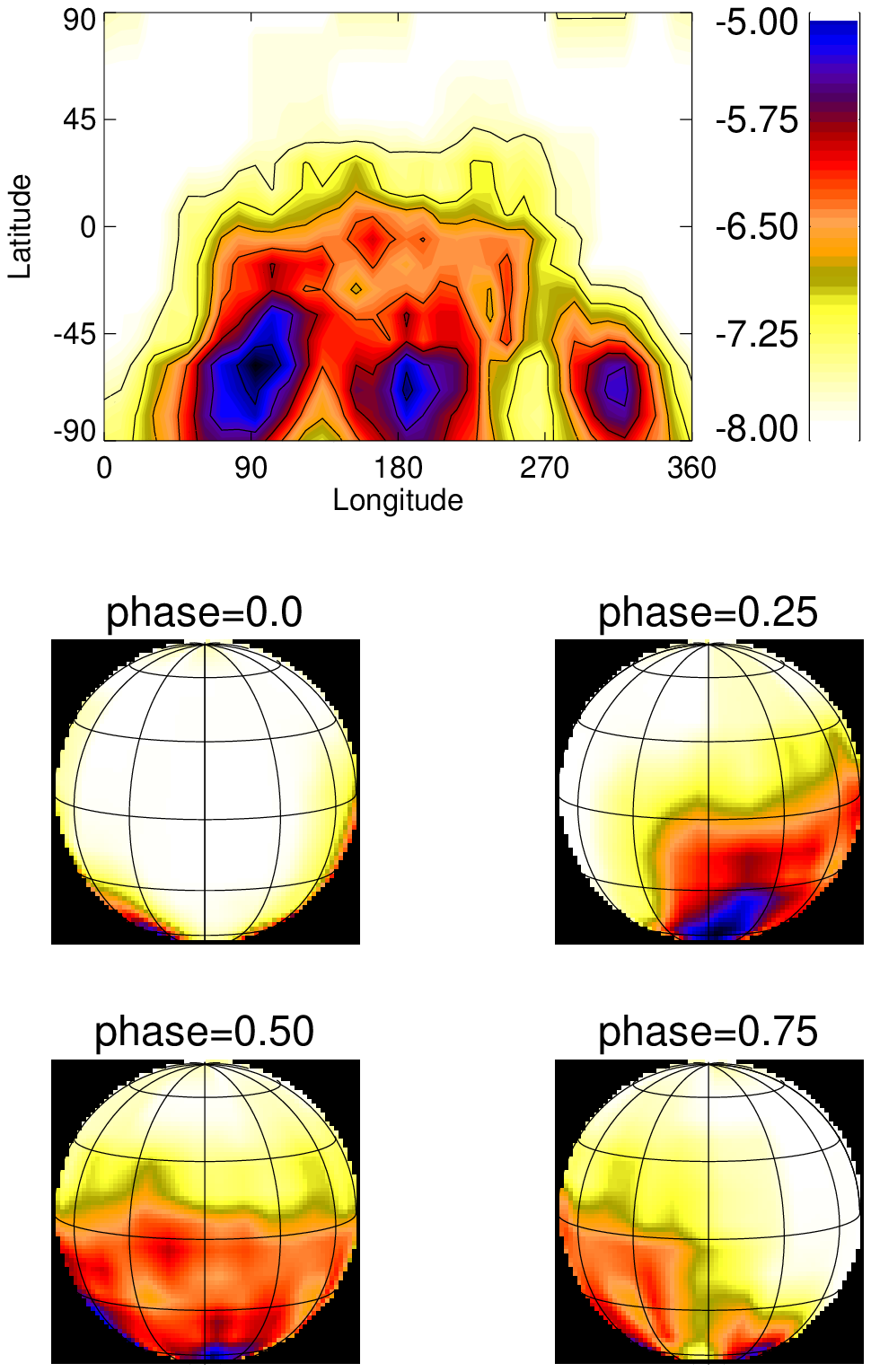}
\caption{
Maps of the abundance distribution for Fe (left), Sr (middle), and Y (right)
on the surface of the primary in the system AR Aur.
-- Credit: Hubrig et al., A\&A, 547, A90, 2012, reproduced with permission \copyright ESO.
}
\label{fig:ARAur_maps}
\end{figure}

\begin{figure}[t]
\centering
\includegraphics[width=0.33\textwidth]{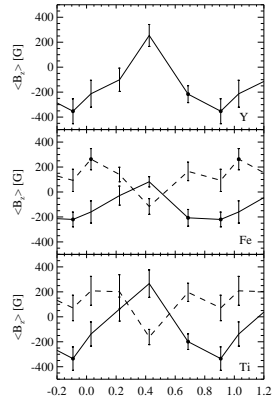}
\caption{
Measurements of the mean longitudinal magnetic field presented as a function of the rotation phase for AR\,Aur.
They were carried out separately for the elements Ti, Fe, and Y (from bottom to top).
The solid line denotes the primary component, while the dashed line denotes the secondary component.
Filled circles indicate 3$\sigma$ measurements.
-- Credit: Hubrig et al., A\&A, 547, A90, 2012, reproduced with permission \copyright ESO.
}
\label{fig:ARAur_magn}
\end{figure}

One of the most exciting objects containing a HgMn star is the triple system AR\,Aur. 
The inner two stars constitute the only known eclipsing binary encompassing a HgMn star.
This binary has an orbital period of 4.13\,d and an age of 4\,Myr.
The two stars are of spectral types B9V and B9.5V, and while the primary HgMn star
is exactly on the ZAMS, the secondary is still contracting
(e.g.\ Nordstr\"om \& Johansen 1994).
Hubrig et al.\ (2012) used observations with SOFIN at the Nordic Optical Telescope
to study the distribution of different elements over the surface of the primary HgMn
star, using the Doppler mapping technique (see Fig.~\ref{fig:ARAur_maps}).
From the same data set, they also determined the magnetic field in both primary and secondary (Fig.~\ref{fig:ARAur_magn}).
AR\,Aur shows a similar behavior to other HgMn systems discussed by Hubrig et al.\ (2012).
The results suggest the existence of a correlation between the magnetic field,
the abundance anomalies, and the binary properties.
For the synchronously rotating components of the SB2 system AR\,Aur, it looks as if
the stellar surfaces facing the companion star usually
display low-abundance element spots and negative magnetic field polarity.
The surface of the opposite hemisphere, as a rule, is covered by high-abundance
element spots and the magnetic field is positive at the rotation phases of
the best-spot visibility (Hubrig et al.\ 2010).
Still, the discussion about the presence of weak magnetic fields in HgMn stars
is still ongoing (see Kochukhov et al.\ 2013).

\section{Summary}
\label{sect:summary}

Chemically peculiar stars are probably the most challenging main sequence stars
to model due to their magnetic fields, and the element segregation and stratification.
They are ideal atomic physics laboratories.
They are the best objects to learn about magnetic field models, to be applied to other classes of stars.
Binarity and multiplicity for the different classes is different,
which potentially leads to new insights into star formation mechanisms.

%
%
%

%
%
%

\end{document}